\documentclass[12pt]{article}
\usepackage{epsfig,amssymb,amsmath,psfrag}

\catcode`\@=11
\DeclareMathAlphabet{\mathpzc}{OT1}{pzc}{m}{it}

\newcommand{\insertfig}[2]{\mbox{\epsfxsize=#1cm \epsfbox{#2.eps}}}

\font\cmss=cmss12 
\def\1{\hbox{{1}\kern-.25em\hbox{l}}}
\def\bfZ{\relax{\hbox{\cmss Z\kern-.4em Z}}}

\def \be  {\begin{equation}}
\def \ee  {\end{equation}}
\def \ba  {\begin{eqnarray}}
\def \ea  {\end{eqnarray}}
\def \baa {\begin{eqnarray*}}
\def \eaa {\end{eqnarray*}}
\def \bb  {\begin {thebibliography} }
\def \eb  {\end{thebibliography}}
\def \lab #1 {\label{#1}}

\newcommand\re[1]{(\ref{#1})}

\def \matrix #1 {\left(\begin{array}{cc} #1 \end{array}\right)}

\def \tr {\mathop{\rm tr}\nolimits}

\newcommand{\as}{\ifmmode\alpha_{\rm s}\else{$\alpha_{\rm s}$}\fi}
\newcommand{\asbar}{\ifmmode\bar{\alpha}_{\rm s}\else{$\bar{\alpha}_{\rm s}$}\fi}

\newcommand{\bit}[1]{\mbox{\boldmath$#1$}}
\newcommand{\ft}[2]{{\textstyle\frac{#1}{#2}}}

\font\cmss=cmss12 
\def\inbar{\,\vrule height1.5ex width.4pt depth0pt}
\def\IC{\relax\hbox{$\inbar\kern-.3em{\rm C}$}}
\def\IZ{\relax{\hbox{\cmss Z\kern-.4em Z}}}
\def\IR{{\hbox{{\rm I}\kern-.2em\hbox{\rm R}}}}

\def\IP{{\hbox{{\rm I}\kern-.2em\hbox{\rm P}}}}
\def\II{\hbox{{1}\kern-.25em\hbox{l}}}

\def\numberbysection{\@addtoreset{equation}{section}
                     \def\theequation{\thesection.\arabic{equation}}}
\numberbysection

\newbox\lett\newdimen\lheight\newdimen\lwidth
\def\ontop#1#2{\setbox\lett=\hbox{#2}\lheight\ht\lett
\multiply\lheight by 12 \divide\lheight by 10\relax%
\lwidth\wd\lett \multiply\lwidth by 8 \divide\lwidth by 10\relax #2\kern-\lwidth%
\raise\lheight\hbox{{$\scriptstyle #1$}}\kern.1ex}

\def\XXint#1#2#3{{\setbox0=\hbox{$#1{#2#3}{\int}$}
     \vcenter{\hbox{$#2#3$}}\kern-.5\wd0}}

\begin{document}

\begin{titlepage}

\thispagestyle{empty}

\vskip2cm

\centerline{\large \bf Fusion hierarchies for $\mathcal{N} = 4$ superYang-Mills theory}

\vspace{1cm}

\centerline{\sc A.V. Belitsky}

\vspace{10mm}

\centerline{\it Department of Physics, Arizona State University}
\centerline{\it Tempe, AZ 85287-1504, USA}

\vspace{3cm}

\centerline{\bf Abstract}

\vspace{5mm}

We employ the analytic Bethe Anzats to construct eigenvalues of transfer matrices with
finite-dimensional atypical representations in the auxiliary space for the putative
long-range spin chain encoding anomalous dimensions of all composite single-trace gauge
invariant operators of the maximally supersymmetric Yang-Mills theory. They obey an
infinite fusion hierarchy which can be reduced to a finite set of integral relations for
a minimal set of transfer matrices. This set is used to derive a finite systems of
functional equations for eigenvalues of nested Baxter polynomials.

\end{titlepage}

\setcounter{footnote} 0

\newpage

\pagestyle{plain}
\setcounter{page} 1

\section{Introduction}

Up until very recent years, only one-dimensional quantum field theories entertained
nonperturbative analytical considerations in coupling constant \cite{Tsv03,Sut04}. The
reason behind availability of these results is largely due to exact solubility of
corresponding models \cite{Bax82,KorBogIze93,Sut04}. Four-dimensional non-supersymmetric
gauge theories were also known to possess integrable structures. In certain limits,
their dynamics effectively becomes low-dimensional and one observes appearance of
structures similar to the interacting spin chains. The high-energy limit of QCD scattering
is described by a two-dimensional reggeon theory \cite{Lip96}, whose quantum mechanical
Hamiltonian was shown to possess holomorphic separability \cite{Lip93} with each component
being identified with the Hamiltonian of the one-dimensional noncompact $sl(2, \mathbb{C})$
Heisenberg magnet \cite{Lip93,FadKor95}. In a different limit, the one-loop renormalization
group evolution of quasipartonic Wilson operators \cite{BukFroKurLip85} is governed by a
pairwise light-cone dilatation operator. In the multi-color limit, only nearest-neighbor
interactions survive in it and its maximal-helicity subsector is mapped to the XXX $sl(2,
\mathbb{R})$ spin chain \cite{BraDerMan98,Bel98,BraDerKorMan99,Bel99}. In both cases, the
models were solved by powerful techniques such as Bethe Ansatz \cite{TakFad79,KorBogIze93}
and Baxter $Q-$operator \cite{Bax72}.

The integrable spin chain underlying the QCD maximal-helicity dilatation operator gets
extended to sectors with other field contents in gauge theories enjoying supersymmetry
\cite{Lip98,MinZar03,BeiKriSta03,Bei03,BeiSta03,BelGorKor03,BelDerKorMan04}. In the
maximally supersymmetric Yang-Mills theory, anomalous dimensions of all Wilson operators
\cite{Lip98,Bei03,BelDerKorMan04} are encoded in the underlying $su(2,2|4)$ spin chain
\cite{BeiSta03} that can be diagonalized by means of the nested Bethe Ansatz \cite{Yan67}.
The symmetry algebra of the magnet reflects the classical symmetry of the gauge theory.
Moreover, the $\mathcal{N} = 4$ supersymmetric quantum field theory, being anomaly-free,
remains superconformal even at the quantum level such that the absence of bound states
in its spectrum warrants anomalous dimensions to be the relevant physical observables. Thus
their calculation implies solution of the field theory in question. What is even more
extraordinary is that in the $\mathcal{N} = 4$ superYang-Mills theory, the one-loop integrability,
manifesting itself through spectral degeneracies, persists beyond leading orders of
perturbation theory. This was demonstrated by explicit multi-loop diagrammatic calculations
in compact \cite{BeiKriSta03} and noncompact \cite{BelKorMul05,Ede05} subsectors of the
theory. This evidence, together with multiloop field-theoretical calculations for twist-two
operators \cite{VogMocVer04,KotLip04}, pointed out that anomalous dimensions of corresponding
gauge invariant operators can be deduced from a kind of a deformed Bethe Ansatz as was shown
in Ref.\ \cite{Sta04}. Together with insights \cite{KazMarMinZar04,BeiDipSta04} into exact
solvability of this gauge theory coming from its conjectured duality to the type IIB string
theory on the AdS$_5 \times$S$^5$ background \cite{Mal97}, a set of all-order nested Bethe
Ansatz equations was suggested in Ref.\ \cite{BeiSta05}.

The conjectured long-range Bethe Ansatz equations are exact only for infinitely long spin
chains. For finite-size systems they become invalid when the interaction range becomes longer
than the number of fields in the gauge-invariant operator. Thus they are only asymptotic
in character (see, e.g., Ref.\ \cite{KotLipRejStaVel07}) and this calls for consistent
incorporation of finite-size effects into the current treatment. One approach to resolve
this problem is known as the Thermodynamics Bethe Ansatz (TBA) \cite{Zam90,BazLukZam97}
extended to lattice systems in Refs.\ \cite{KluPea92,JutKluSuz98}. A pivotal role in the
derivation of TBA equations, which is independent of the notorious string hypothesis, is
played by the functional algebraic relations among the hierarchy of fused row-to-row
higher spin transfer matrices \cite{BazRes89,KunOhtSuz95,KunNakSuz93,Tsu97}. For simplest
cases, these relations were shown to have a deep mathematical origin in exact sequence of
Yangian modules \cite{KunNakSuz93}. The resulting fusion hierarchy yields an infinite system
of coupled equations which can be consistently truncated only for finite temperature and is
hard to practically use otherwise for vanishing temperatures \cite{Tsu02}. However, it was
demonstrated by Takahashi \cite{Tak01}, and extended to superspin chains by Tsuboi \cite{Tsu02},
that the infinite hierarchy of TBA equations can be reduced to a finite set of integral
equations for a finite number of fused transfer matrices. This represent an alternative
and more practical approach to study the thermodynamics of integrable models.

In this paper we address the question of constructing fusion hierarchies among transfer
matrices for the putative $su(2,2|4)$ long-range spin chain of the maximally supersymmetric
Yang-Mills theory and their use for the solution of the spectral problem. The peculiarities
of the model exposed in renormalization of the spectral parameter and various dressing factors
make the analytical properties of fused transfer matrices quite complicated to allow for
generalization of Takahashi equations. Consequently, we focus on the derivation of closed
systems of Baxter equations involving minimal sets of fused transfer matrices. These can be
diagonalized either order-by-order of perturbation theory \cite{BelKorMul06,Bel06,Bel07a,Bel07b}
or used for strong coupling analyses as was demonstrated earlier for the long-range rank-one
$sl(2)$ spin chain \cite{Bel07c}.

\section{Analytic Bethe Ansatz}
\label{AnalyticBA}

The main ingredient of the current framework are row-to-row transfer matrices that encode
full set of mutually commuting conserved charges of the chain. For short-range spin chains,
the $R-$matrix formalism provides the relevant framework for systematic construction of
commuting transfer matrices. We possess however only the conjectured nested Bethe Ansatz
equations \cite{BeiSta05} for the underlying long-range magnet. The lack of a systematic
microscopic approach does not prevent one from finding the eigenvalues of fused transfer
matrices though. The main r\^ole in this situation is played by the analytic Bethe Ansatz
\cite{Res83} which relies on general macroscopic properties of the system such as analyticity.

\begin{figure}[t]
\begin{center}
\mbox{
\begin{picture}(0,40)(200,0)
\put(0,0){\insertfig{14}{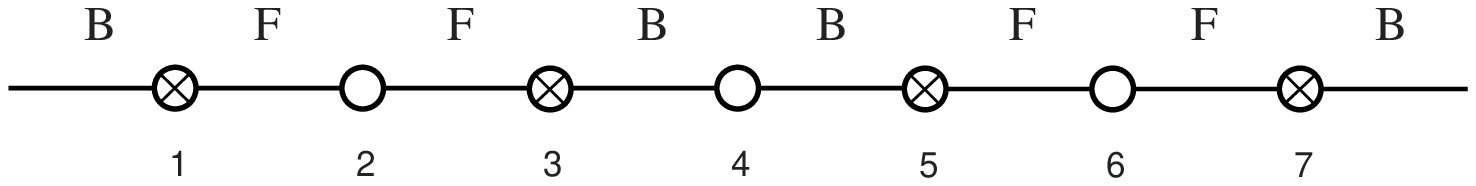}}
\end{picture}
}
\end{center}
\caption{\label{Dynkin} Kac-Dynkin diagram used in the construction of transfer
matrices.}
\end{figure}

For the spin chain based on the high-rank $su(2,2|4)$ superalgebra, there are several
equivalent ways to formulate the nested Bethe Ansatz \cite{Kul86,ResWie87}. They reflect
different choices of its simple root system $\{ \bit{\alpha}_p |p = 1, {\dots}, 7 \}$
\cite{Kac77}. For the rational short-range $R-$matrix, which generates the spectrum of
one-loop anomalous dimensions, they take the following generic form
\be
\label{NestedABAgeneric}
(- 1)^{ A_{pp}/2} \left(
\frac{u^{(p)}_{0,k} - \ft{i}{2} w_p}{u^{(p)}_{0,k} + \ft{i}{2} w_p}
\right)^L
=
\prod_{q = 1}^{7} \prod_{j = 1}^{n_q}
\frac{
u^{(p)}_{0,k} - u^{(q)}_{0,j} + \ft{i}{2} A_{pq}
}{
u^{(p)}_{0,k} - u^{(q)}_{0,j} - \ft{i}{2} A_{pq}
}
\, ,
\ee
with the structure of the algebra encoded in the Cartan matrix $A_{pq} = (\bit{\alpha}_p |
\bit{\alpha}_q)$ and the representation on the spin-chain sites determined by the Kac-Dynkin
labels, $w_p = (\bit{\alpha}_p| \bit{\mu})$ expressed by the corresponding weight vector
$\bit{\mu}$. However, out of many equivalent Kac-Dynkin diagrams only four allow for an
all-order generalization \cite{BeiSta05}, i.e., consistent deformation of \re{NestedABAgeneric}
with the 't Hooft coupling $g = g_{\rm\scriptscriptstyle YM} \sqrt{N_c}/(2 \pi)$. They
correspond to the ground state $| \Omega \rangle$ built from the complex scalar $Z-$fields
of the theory, $| \Omega \rangle = \tr[ Z^L (0) ]$, which is protected from quantum
corrections to all order of perturbation theory. We choose to develop our formalism based
on the Kac-Dynkin diagram displayed in Fig.\ \ref{Dynkin} with its central node corresponding
to the light-cone $sl(2)$ subalgebra formed by projections of the generators of the Lorentz
transformations, translations and conformal boosts,
\be
i \mathfrak{M}^{-+} = \mathbb{L}_1{}^1 + \bar{\mathbb{L}}^{\dot{1}}{}_{\dot{1}}
\, , \qquad
i \mathfrak{P}^+ = \sqrt{2} \mathbb{P}_{1\dot{1}}
\, , \qquad
i \mathfrak{K}^- = \sqrt{2} \mathbb{K}^{\dot{1}1}
\, ,
\ee
respectively. The pseudovacuum of the second level Bethe Ansatz corresponds the
derivative sector $| \Omega^\prime \rangle = \tr[ \mathcal{D}_+^N Z^L ]$ universal to all
gauge theories. The only nontrivial Kac-Dynkin label of the gauge-theory spin chain is
associated with the central node, $w_4 = - 1$. The complete Serre-Chevalley basis is
spelled out in Appendix \ref{SuperspaceRealization}.

Transfer matrices are supertraces of monodromy matrices with a chosen representation of
the symmetry algebra in the auxiliary space. Thus their eigenvalue formulas are expected
to be a sum of terms corresponding to each components of a given representation labelled
by the Young diagram as was demonstrated in Refs.\ \cite{BazRes89,KunOhtSuz95} and
\cite{Tsu97} for classical and graded algebras, respectively. Thus the construction is
based on the identification of the elementary Young supertableaux which depend on the
spectral parameter and read for the long-range $su(2,2|4)$ magnet\footnote{See Appendix
\ref{su2Dynkin} for Young tableaux corresponding to the Kac-Dynkin diagram with the
FBBFFBBF grading, i.e., with the central compact $su(2)$ node.}
\ba
\label{DistinguishedBox}
\unitlength0.4cm
\begin{picture}(2.3,1.2)
\linethickness{0.4mm}
\put(1,0){\framebox(1,1){$1$}}
\end{picture}
{}_{u}
\!\!\!&=&\!\!\!
\left( x^+ \right)^L
\frac{
\widehat{Q}^{(1)} \left( u^+ \right)}{\widehat{Q}^{(1)} \left( u^- \right)}
{\rm e}^{\ft12
\left[
\Delta_+ (u^-) - \Delta_- (u^-) + \Delta_+ (u^+) + \Delta_- (u^+)
+ \sigma_0^{(3)} (u^-) - \sigma_0^{(3)} (u^+)
\right]}
\, , \\
\unitlength0.4cm
\begin{picture}(2.3,1)
\linethickness{0.4mm}
\put(1,0){\framebox(1,1){$2$}}
\end{picture}
{}_{u}
\!\!\!&=&\!\!\!
\left( x^+ \right)^L
\frac{
\widehat{Q}^{(1)} \left( u^+ \right)}{\widehat{Q}^{(1)} \left( u^- \right)}
\frac{
Q^{(2)} \left( u - i \right)}{Q^{(2)} \left( u \right)}
{\rm e}^{\ft12
\left[
\Delta_+ (u^+) + \Delta_- (u^+) + \sigma_0^{(3)} (u^-) - \sigma_0^{(3)} (u^+)
\right]}
\, , \nonumber\\
\unitlength0.4cm
\begin{picture}(2.3,1)
\linethickness{0.4mm}
\put(1,0){\framebox(1,1){$3$}}
\end{picture}
{}_{u}
\!\!\!&=&\!\!\!
\left( x^+ \right)^L
\frac{
Q^{(2)} \left( u + i \right)}{Q^{(2)} \left( u \right)}
\frac{
\widehat{Q}^{(3)} \left( u^- \right)}{\widehat{Q}^{(3)} \left( u^+ \right)}
{\rm e}^{\ft12
\left[
\Delta_+ (u^+) + \Delta_- (u^+) + \sigma_0^{(1)} (u^+) - \sigma_0^{(1)} (u^-)
\right]}
\, , \nonumber\\
\unitlength0.4cm
\begin{picture}(2.3,1)
\linethickness{0.4mm}
\put(1,0){\framebox(1,1){$4$}}
\end{picture}
{}_{u}
\!\!\!&=&\!\!\!
\left( x^+ \right)^L
\frac{
\widehat{Q}^{(3)} \left( u^- \right)}{\widehat{Q}^{(3)} \left( u^+ \right)}
\frac{
Q^{(4)} \left( u + i \right)}{Q^{(4)} \left( u \right)}
{\rm e}^{ \Delta_+ (u^+)
+ \ft12 \left[ \sigma_0^{(1)} (u^+) - \sigma_0^{(1)} (u^-) \right]}
\, , \nonumber\\
\unitlength0.4cm
\begin{picture}(2.3,1)
\linethickness{0.4mm}
\put(1,0){\framebox(1,1){$5$}}
\end{picture}
{}_{u}
\!\!\!&=&\!\!\!
\left( x^- \right)^L
\frac{
\widehat{Q}^{(5)} \left( u^+ \right)}{\widehat{Q}^{(5)} \left( u^- \right)}
\frac{
Q^{(4)} \left( u - i \right)}{Q^{(4)} \left( u \right)}
{\rm e}^{ \Delta_- (u^-)
+ \ft12 \left[ \sigma_0^{(7)} (u^-) - \sigma_0^{(7)} (u^+) \right]}
\, , \nonumber\\
\unitlength0.4cm
\begin{picture}(2.3,1)
\linethickness{0.4mm}
\put(1,0){\framebox(1,1){$6$}}
\end{picture}
{}_{u}
\!\!\!&=&\!\!\!
\left( x^- \right)^L
\frac{
Q^{(6)} \left( u - i \right)}{Q^{(6)} \left( u \right)}
\frac{
\widehat{Q}^{(5)} \left( u^+ \right)}{\widehat{Q}^{(5)} \left( u^- \right)}
{\rm e}^{\ft12
\left[
\Delta_- (u^-) + \Delta_+ (u^-) + \sigma_0^{(7)} (u^-) - \sigma_0^{(7)} (u^+)
\right]}
\, , \nonumber\\
\unitlength0.4cm
\begin{picture}(2.3,1)
\linethickness{0.4mm}
\put(1,0){\framebox(1,1){$7$}}
\end{picture}
{}_{u}
\!\!\!&=&\!\!\!
\left( x^- \right)^L
\frac{
\widehat{Q}^{(7)} \left( u^- \right)}{\widehat{Q}^{(7)} \left( u^+ \right)}
\frac{
Q^{(6)} \left( u + i \right)}{Q^{(6)} \left( u \right)}
{\rm e}^{\ft12
\left[
\Delta_- (u^-) + \Delta_+ (u^-) + \sigma_0^{(5)} (u^+) - \sigma_0^{(5)} (u^-)
\right]}
\, , \nonumber\\
\unitlength0.4cm
\begin{picture}(2.3,1.2)
\linethickness{0.4mm}
\put(1,0){\framebox(1,1){$8$}}
\end{picture}
{}_{u}
\!\!\!&=&\!\!\!
\left( x^- \right)^L
\frac{
\widehat{Q}^{(7)} \left( u^- \right)}{\widehat{Q}^{(7)} \left( u^+ \right)}
{\rm e}^{\ft12
\left[
\Delta_- (u^+) - \Delta_+ (u^+) + \Delta_- (u^-) + \Delta_+ (u^-)
+ \sigma_0^{(5)} (u^+) - \sigma_0^{(5)} (u^-)
\right]}
\, .
\ea
Here each box corresponds to the respective bond connecting the nodes of the graph in Fig.\
\ref{Dynkin} with assigned bosonic $\bar{1} = \bar{4} = \bar{5} = \bar{8} = 0$ and fermionic
$\bar{2} = \bar{3} = \bar{6} = \bar{7} = 1$ gradings. These tableaux are built from bare and
dressed Baxter polynomials,
\be
Q^{(p)} (u) = \prod_{k = 1}^{n_p} \left( u - u^{(p)}_k \right)
\, , \qquad
\widehat{Q}^{(p)} (u) = \prod_{k = 1}^{n_p} \left( x - x^{(p)}_k \right)
\, ,
\ee
expressed in terms of, respectively, bare $u$ and renormalized spectral parameter $x = x [u]
\equiv \ft12 \big( u + \sqrt{u^2 - g^2} \big)$ \cite{BeiDipSta04}, and assumed shorthand
notations $u^\pm \equiv u \pm \ft{i}{2}$ and $x^\pm \equiv x [u^\pm]$. The Bethe roots
$u^{(p)}_k$ admit an infinite series expansion in the 't Hooft coupling, $ u^{(p)}_k  =
u^{(p)}_k (g) =  u^{(p)}_{0,k} + g^2  u^{(p)}_{1,k} + \mathcal{O} (g^4)$. The exponential
factors correspond to dressing factors which emerge at higher orders of perturbation theory
and read
\be
\label{Sigmap}
\sigma^{(p)}_\eta (u) = \int_{-1}^1 \frac{d t}{\pi}
\frac{\ln Q^{(p)} (\eta \ft{i}{2} - g t)}{\sqrt{1 - t^2}}
\left(
1 - \frac{\sqrt{u^2 - g^2}}{u + g t}
\right)
\, ,
\ee
where $\eta$ takes three possible values $\pm, 0$. For $\eta = 0$, it can be represented
as a ratio of the bare and dressed Baxter polynomials
\be
\frac12 \sigma_0^{(p)} (u^\pm)
=
\ln \frac{\widehat{Q}^{(p)} (u^\pm)}{Q^{(p)} (u^\pm)}
\, .
\ee
For the momentum carrying Bethe roots, there is an additional contribution $\Theta (u)$
\be
\Delta_\pm (u) = \sigma_\pm^{(4)} (u) - \Theta (u)
\, ,
\ee
known as the magnon dressing phase which is responsible for smooth interpolation between
the weak- and strong-coupling expansions \cite{AruFroSta04,BeiHerLop06,BeiEdeSta06} and
reads in terms of the Baxter polynomials \cite{Bel07a}
\ba
\label{ThetaWeak}
\Theta (u)
\!\!\!&=&\!\!\!
g \int_{- 1}^1 \frac{d t}{\sqrt{1 - t^2}}
\ln \frac{Q^{(4)} ( - \ft{i}{2} - g t )}{Q^{(4)} ( + \ft{i}{2} - g t )}
\,
{-\!\!\!\!\!\!\int}_{-1}^1 ds \frac{\sqrt{1 - s^2}}{s - t}
\nonumber\\
&\times&
\int_{C_{[i, i \infty]}} \frac{d \kappa}{2 \pi i} \frac{1}{\sinh^2 (\pi \kappa)} \
\ln
\left( 1 + \frac{g^2}{4 x x [\kappa + g s]} \right)
\left( 1 - \frac{g^2}{4 x x [\kappa - g s]} \right)
\, .
\ea

Introducing a symbolic notation for the differential operator involving the single-box
Young supertableau introduced above, labelled by the flavor index $\alpha$ and the
spectral parameter $u$,
\be
\mathcal{D}_{\alpha, [\pm]}
\equiv
1 \pm \!\!\!\!
{\unitlength0.4cm
\begin{picture}(2.3,1.2)
\linethickness{0.4mm}
\put(1,-0.2){\framebox(1,1){$\alpha$}}
\end{picture}}
{}_{u} \,
{\rm e}^{-i \partial_u}
\, ,
\ee
one can write the generating functions \cite{BazRes89,KunOhtSuz95,KriLipWieZab97,Tsu97}
for the eigenvalues of transfer matrices with atypical finite-dimensional antisymmetric
$(1^a)$ and symmetric $(s)$ representations in the auxiliary space,
\ba
\label{TransferAntiSym}
\mathcal{D}_{8, [+]}
\mathcal{D}_{7, [+]}^{-1}
\mathcal{D}_{6, [+]}^{-1}
\mathcal{D}_{5, [+]}
\mathcal{D}_{4, [+]}
\mathcal{D}_{3, [+]}^{-1}
\mathcal{D}_{2, [+]}^{-1}
\mathcal{D}_{1, [+]}
\!\!\!&=&\!\!\!
\sum_{a = 0}^\infty
t_{[a]} \left( u^{[1 - a]} \right) {\rm e}^{- i a \partial_u}
\, , \\
\label{TransferSym}
\mathcal{D}_{1, [-]}^{-1}
\mathcal{D}_{2, [-]}
\mathcal{D}_{3, [-]}
\mathcal{D}_{4, [-]}^{-1}
\mathcal{D}_{5, [-]}^{-1}
\mathcal{D}_{6, [-]}
\mathcal{D}_{7, [-]}
\mathcal{D}_{8, [-]}^{-1}
\!\!\!&=&\!\!\!
\sum_{s = 0}^\infty
t^{\{s\}} \left( u^{[1 - s]} \right) {\rm e}^{- i s \partial_u}
\, ,
\ea
where
\be
u^{[\pm n]} \equiv u \pm \ft{i}{2} n
\, , \qquad
x^{[\pm n]} \equiv x [u^{[\pm n]}]
\, ,
\ee
which are generalizations of the previously introduced conventions $u^{[\pm 1]} = u^\pm$.
In Eqs.\ \re{TransferAntiSym} and \re{TransferSym}, the powers $p_\alpha = 1 - 2 \bar{\alpha}$
of the multiplicative factors in the left-hand side of the equations reflect the grading of
the bonds of the Kac-Dynkin diagram. The residues of apparent poles positioned at zeroes of
the Baxter polynomials in these transfer matrices vanish upon the use of the long-range
nested Bethe Ansatz equations of Ref.\ \cite{BeiSta05}.

The momentum carrying Baxter polynomial $Q^{(4)} (u)$ determines the anomalous dimensions
\be
\label{AllOrderAD}
\gamma (g) = i g^2 \int_{- 1}^1 \frac{dt}{\pi} \sqrt{1 - t^2}
\left(
\ln \frac{
Q^{(4)} \left( + \ft{i}{2} - g t \right)
}{
Q^{(4)} \left( - \ft{i}{2} - g t \right)
}
\right)^\prime
\, ,
\ee
of local Wilson operators with quantum numbers defined by the labels of the Kac-Dynkin
diagram, Fig.\ \ref{Dynkin}, as demonstrated in Appendix \ref{SuperspaceRealization}.
Solutions corresponding to the anomalous dimensions of Wilson operators are selected
by the vanishing quasimomentum $\theta$ condition imposed on the states of the chain
\be
i \theta
=
\frac{1}{\pi}
\int_{- 1}^1 \frac{dt}{\sqrt{1 - t^2}}
\ln \frac{Q^{(4)} (+ \ft{i}{2} - g t)}{Q^{(4)} (- \ft{i}{2} - g t)}
= 0
\, .
\ee

For the $su(2,2|4)$ magnet, the strategy to determine the eigenvalues $Q^{(4)} (u)$ consists
in selecting the first seven fused transfer matrices, either $t_{[a]} (u)$ or $t^{\{s\}} (u)$,
and subsequent solution of the resulting system of equations for the zeroes of transfer
matrices and Baxter polynomials. For the Kac-Dynkin diagram we have chosen for our analysis,
one immediately finds that $t_{[a]}(u)$ possess kinematical zeroes
\be
t_{[a]} (u)
=
\prod_{j = 0}^{a - 2} \left( x^{[a - 2 - 2j]} \right)^L
T_{[a]} (u)
\, ,
\ee
contrary to the ones with symmetric auxiliary space $t^{\{s\}} (u)$. Therefore, while for the
former one has sufficient number of equations to determine all unknowns, it is no longer the
case for the symmetric transfer matrices that contain a redundant set of conserved charges.
The latter are not completely independent though and can be related by means of nonlinear
integral equations obeyed by $t^{\{s\}} (u)$ \cite{Tak01,Tsu02}, as we demonstrate at one
loop in Section \ref{TakahahiEquations}. We will study the diagonalization procedure below
on examples of low-rank subsectors. It will turn out that due to simplicity of corresponding
fused transfer matrices, the rank-two $sl(2|1)$ subsector allows for complete resolution of
nesting. Above it, the nesting is intrinsic, however, it may disappear in specific cases once
the consideration is restricted to the lowest order of perturbation theory.

\section{T-system}
\label{Tsystem}

The antisymmetric and symmetric transfer matrices \re{TransferAntiSym} and \re{TransferSym}
discussed in the previous Section obey an infinite set of algebraic functional relations.
To find them, we introduce at first transfer matrices with the auxiliary space labelled by
a general Young superdiagram\footnote{Our analysis  can be extended to skew Young supertableaux
as was done in Ref.\ \cite{Bel07b} for the $sl(2|2)$ sector, however, since these will not be
relevant for our subsequent considerations and they will not be presently addressed.}
$Y (\bit{s})$ defined by the partitioning $\bit{s} = \{ s_1, s_2, \dots \}$ with $s_1 \geq
s_2 \geq \dots$ \cite{BazRes89}. To this end, one labels all boxes of the Young superdiagram
$Y (\bit{s})$, starting with the one in the top left corner, with a pair of integers $(j,k)$,
corresponding to rows and columns, respectively. Then one defines a set of admissible Young
supertableaux $Y_\alpha (\bit{s})$ by assigning one of eight $\alpha (j,k)$ ``flavor'' indices
to each box of the diagram $Y (\bit{s})$ and distributing them according to the ordering
conditions: $\alpha (j, k) < \alpha (j, k + 1)$ and $\alpha (j, k) < \alpha (j + 1, k)$. These
strong ordering conditions turn into weaker inequalities when the indices on adjacent boxes
have coincident gradings, namely, for
\begin{itemize}
\item bosonic grading $\bar\alpha = 0$:
\be
\alpha (j, k) \leq \alpha (j, k + 1)
\, , \qquad
\alpha (j, k) < \alpha (j + 1, k)
\, ;
\ee
\item fermionic grading $\bar\alpha = 1$:
\be
\alpha (j, k) < \alpha (j, k + 1)
\, , \qquad
\alpha (j, k) \leq \alpha (j + 1, k)
\, .
\ee
\end{itemize}
The transfer matrix for a rectangular $\bit{s} = \{ s^a \}$ Young superdiagram $Y (\bit{s})$
can be easily constructed according to this set of rules and reads
\be
t_{[a]}^{\{s\}} (u)
=
\sum_{Y_\alpha} \sum_{\alpha (j, k) \in Y_\alpha}
p_{\alpha (j, k)}
{\unitlength0.4cm
\begin{picture}(3,2)
\linethickness{0.4mm}
\put(0.5,-0.7){\framebox(2,2){$\scriptstyle\alpha (j, k)$}}
\end{picture}}
{}_{u^{[s - a + 2j - 2k]}}
\, .
\ee
Obviously, by construction, $t^{\{1\}}_{[a]} (u) = t_{[a]} (u)$ and $t^{\{s\}}_{[1]} (u) =
t^{\{s\}} (u)$. This general rectangular transfer matrix $t_{[a]}^{\{s\}} (u)$ admits the
Bazhanov-Reshetikhin determinant representation \cite{BazRes89,Tsu97} (see also Ref.\
\cite{KazVie07} for a recent discussion) in their terms,
\be
\label{BazhanovReshetikhin}
t_{[a]}^{\{s\}} (u)
=
\det_{1 \leq j, k \leq s} t_{[a - j + k]} (u^{[s + 1 - j - k]})
=
\det_{1 \leq j, k \leq a} t^{\{s + j - k\}} (u^{[j + k - a - 1]})
\, ,
\ee
with boundary conditions $t^{\{n\}} (u) = t_{[n]} (u) = 0$ for $n < 0$. Making use of the
Desnanot-Jacobi identity for determinants, one finds the $T-$system of equations
\cite{KluPea92,KunNakSuz93,KriLipWieZab97,Tsu97} (see also \cite{KazSorZab07}) obeyed by
rectangular transfer matrices,
\ba
t_{[a]}^{\{s\}} (u^+) t_{[a]}^{\{s\}} (u^-)
=
t_{[a]}^{\{s + 1\}} (u) t_{[a]}^{\{s - 1\}} (u)
+
t_{[a + 1]}^{\{s\}} (u) t_{[a - 1]}^{\{s\}} (u)
\, ,
\ea
with the following truncation properties which will be crucial for the analysis
in the following Section
\be
t^{\{s\}}_{[a]} (u) = 0 \, , \qquad \mbox{for} \qquad a, s \geq \mathcal{N} + 1 \, ,
\ee
and the duality relation\footnote{We kept $\mathcal{N}$ arbitrary so that these
results could be applicable to other $sl(\mathcal{N} | \mathcal{N})$ algebras.}
\be
\label{DualityRelation}
t^{\{ \mathcal{N} + n \}}_{[\mathcal{N}]} (u)
=
t^{\{ \mathcal{N} \}}_{[\mathcal{N} + n]} (u)
\, ,
\ee
with $\mathcal{N} = 4$ for $su(2,2|4)$ and $n$ being non-negative integer.

\section{Takahashi integral equations}
\label{TakahahiEquations}

As we pointed out at the end of Section \ref{AnalyticBA}, the system of seven equations
for the Baxter polynomials $Q^{(p)} (u)$ involving the symmetric transfer matrices
$t^{\{s\}} (u)$ yields an underdetermined system of equations for their zeroes. To
reduce the number of unknowns, they have to be supplemented by relations between
redundant conserved charges. The starting point in the derivation of nonlinear equations,
the so-called Takahashi equations, for the symmetric transfer matrices is the fusion
relation
\ba
\label{ExactFusionT}
\prod_{j = 0}^{s - 1} \left( x^{[s - 1 - 2j]} \right)^{- L}
\left(
t^{\{s\}} (u^+) t^{\{s\}} (u^-)
-
t^{\{s + 1\}} (u) t^{\{s - 1\}} (u)
\right)
=
T_{[2]}^{\{s\}} (u)
\, ,
\ea
where we have used the following factorization property of the fused transfer matrix
$t_{[2]}^{\{s\}} (u)$ in its right-hand side
\be
\prod_{j = 0}^{s - 1} \left( x^{[s - 1 - 2j]} \right)^{- L}
t_{[2]}^{\{s\}} (u)
=
T_{[2]}^{\{s\}} (u)
\, ,
\ee
which enables the very possibility to derive nonlinear equations. An obvious way to
relate the charges entering the expansion of transfer matrices is by using the fact
that the renormalized transfer matrix $T_{[2]}^{\{s\}} (u)$ is a polynomial in the
renormalized spectral parameter, so that the residues of poles in the left-hand side
have to vanish on their own order by order in perturbation theory. At leading order,
this can be summarized in a closed set of nonlinear integral equations as we demonstrate
below.

\begin{figure}[t]
\begin{center}
\mbox{
\begin{picture}(0,205)(80,0)
\put(0,0){\insertfig{5}{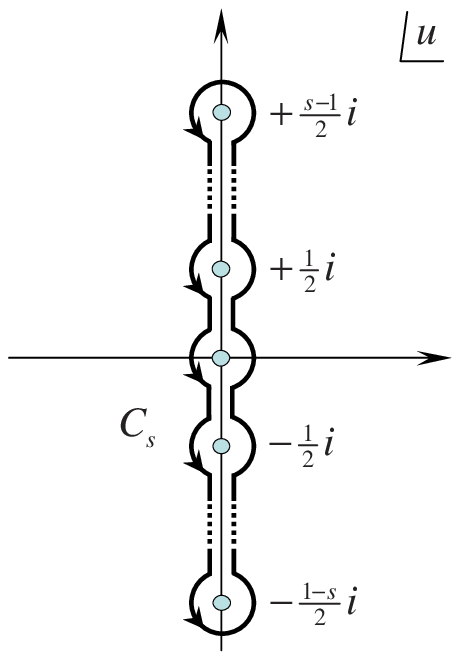}}
\end{picture}
}
\end{center}
\caption{\label{contour} The integration contour in Eq.\ \re{IntegralRep}.}
\end{figure}

One notices that the fusion hierarchy \re{ExactFusionT} can be rewritten as
\ba
\label{FusionT1}
\prod_{j = 0}^{s - 1} \left( u^{[s - 1 - 2j]} \right)^{- L}
\left(
t_0^{\{s\}} (u^+)
-
\frac{
t_0^{\{s + 1\}} (u) t_0^{\{s - 1\}} (u)
}
{
t_0^{\{s\}} (u^-)
}
\right)
=
\frac{
T_{0, [2]}^{\{s\}} (u)
}
{
t_0^{\{s\}} (u^-)
}
\, .
\ea
The crucial observation is that the left-hand side of this equation possesses poles (at
most of order $L$) at $u = - \ft{i}2 (s - 1 - 2 j)$ with $j = 0, \dots, s - 1$, while
the right-hand side does not. Expanding the first term in the left-hand side as a sum of
poles,
\be
\prod_{j = 0}^{s - 1} \left( u^{[s - 1 - 2j]} \right)^{- L} t_0^{\{s\}} (u^+)
=
\sum_{j = 0}^{s - 1} \sum_{k = 1}^L a_j (k) \left( u^{[s - 1 - 2j]} \right)^{-k}
\, ,
\ee
one finds the coefficients $a_j (k)$ by integrating both sides of this equation in the
vicinity of singularities. Here we used the fact that the coefficient of the leading
asymptotic term at large $u$ vanishes, $t^{\{s\}} (u) = 0 \cdot u^{s L} + \mathcal{O}
(u^{sL - 1})$. Performing the summation over $k$, one deduces
\be
\label{IntegralRep}
\prod_{j = 0}^{s - 1} \left( u^{[s - 1 - 2j]} \right)^{- L} t_0^{\{s\}} (u^+)
=
\oint_{C_s} \frac{d v}{2 \pi i} \frac{1}{u - v}
\prod_{j = 0}^{s - 1} \left( v^{[s - 1 - 2j]} \right)^{- L}
t_0^{\{s\}} (v^+)
\, ,
\ee
where the integration contour $C_s$ is shown in Fig.\ \ref{contour}. Substituting \re{FusionT1}
into the above equation, we can neglect its right-hand side since it does not have poles at $u = -
\ft{i}2 (s - 1 - 2 j)$ and obtain a system of integral equation for eigenvalues of transfer
matrices
\be
t_0^{\{s\}} (u^+)
=
\oint_{C_s} \frac{d v}{2 \pi i} \frac{1}{u - v}
\prod_{j = 0}^{s - 1}
\left(
\frac{
u^{[s - 1 - 2j]}
}{
v^{[s - 1 - 2j]}
} \right)^L
\frac{
t_0^{\{s + 1\}} (v) t_0^{\{s - 1\}} (v)
}
{
t_0^{\{s\}} (v^-)
}
\, .
\ee
As it stands, this is an infinite set of nonlinear integral equations. However, as a consequence
of the identity \re{DualityRelation}, it truncates. Namely, using the Bazhanov-Reshetikhin formula
for both sides of this equality, we can solve it for $t^{\{2 \mathcal{N} \}}_0 (u)$ and find the
latter in terms of the transfer matrices $t^{\{s\}}_0 (u)$ with $1 \leq s < 2 \mathcal{N}$,
\ba
\label{SolutionDualityRel}
&&\!\!\!\!\!\!\!\!
t^{\{2 \mathcal{N} \}}_0 (u)
\\
&&\!\!\!\!\!\!\!\!
=
\frac{
\det\limits_{1 \leq j, k \leq \mathcal{N}}
\left( 1 - \delta_{j, \mathcal{N}} \delta_{k,1} \right)
t_0^{ \{ j - k + \mathcal{N} + 1\}} (u^{[j + k - \mathcal{N} - 1]})
-
\det\limits_{1 \leq j, k \leq \mathcal{N} + 1}
\left( 1 - \delta_{j, \mathcal{N} + 1} \delta_{k,1} \right)
t_0^{ \{ j - k + \mathcal{N} \}} (u^{[j + k - \mathcal{N} - 2]})
}{
\det\limits_{1 \leq j, k \leq \mathcal{N} - 1}
t_0^{ \{ j - k + \mathcal{N} \}} (u^{[j + k - \mathcal{N}]})
+
\det\limits_{1 \leq j, k \leq \mathcal{N}}
t_0^{ \{ j - k + \mathcal{N} - 1\}} (u^{[j + k - \mathcal{N} - 1]})
}
. \nonumber
\ea
Such that one gets a finite system of integral equations involving only transfer matrices
$t_0^{\{s\}}$, $1 \leq s \leq 2 \mathcal{N} - 1$ with $\mathcal{N} = 4$. Having reduced
the number of conserved charges, the system of equations for eigenvalues $Q^{(p)} (u)$ can
be solved.

\section{Baxter equations}

In this Section we derive closed systems of TQ-relations, known as Baxter equations,
for Baxter polynomials in low-rank subsectors of the $su(2,2|4)$ long-range spin chain.
Contrary to previous studies \cite{Bel07a,Bel07b} instead of dealing with conjugate
transfer matrices, we will use fused transfer matrices with increasing spin in the
auxiliary space.

First of all, removing all excitations except the ones associated with the central
noncompact node of the Dynkin diagram, we immediately find that the transfer matrix
$t^{\{1\}} (u) = t_{[1]} (u)$ yields the known Baxter equation of the long-range $sl(2)$
magnet \cite{BelKorMul06,Bel07a}. The ground state of this chain is built from the
product of complex scalar fields corresponding to $Z = \bar{\phi}_{43}$ in conventions
of Ref.\ \cite{BelDerKorMan03}, and the light-cone covariant derivatives as elementary
excitations propagating on it. The iterative numerical solution order-by-order in
perturbation theory is given in Appendix \ref{SolutBaxter}. Let us proceed to higher
rank sectors.

\subsection{$sl(2|1)$ sector}

Extending considerations to the case of nonvanishing excitations on the
first fermionic node adjacent to the central one, either left or right, we
obtain the closed $sl(2|1)$ sector. Now in addition to covariant derivatives,
it possesses fermionic excitations, which correspond to either helicity up
$\lambda^1$ or helicity down $\bar\lambda_4$ gauginos, depending on the choice
of the fermionic node. Let us address the case with the nontrivial nodes three
and four, thus $Q^{(p)} (u) = 1$ for $p \neq 3,4$.

One can derive Baxter equations either using symmetric or antisymmetric transfer
matrices. Using the two lowest-dimensional representations in the auxiliary space,
one can eliminate the nested Baxter polynomial $\widehat{Q}^{(3)}$. The resulting
equation for the momentum carrying Baxter polynomial $Q^{(4)} (u)$ remains of the
second order in finite differences, however, it depends quadratically on transfer
matrices. The functional algebraic equation in terms of the symmetric transfer
matrices then reads
\ba
\label{BaxterSymQ4}
\Big( t^{\{2\}} (u^+) \!\!\!&-&\!\!\! t^{\{1\}} (u) t^{\{1\}} (u + i) \Big)
{\rm e}^{\ft12 \Delta_+ (u^{+})}
Q^{(4)} (u + i)
\\
&+&\!\!
(x^{-})^L
\left( t^{\{1\}} (u + i) -  (x^{+})^L {\rm e}^{\ft12 \Delta_- (u^{+}) + \ft12 \Delta_+ (u^{+})} \right)
{\rm e}^{\ft12 \Delta_- (u^{+}) + \Delta_- (u^{-})} Q^{(4)} (u - i)
\nonumber\\
&&
- \,
\left( t^{\{2\}} (u^+) -  (x^{+})^L t^{\{1\}} (u) {\rm e}^{\ft12 \Delta_- (u^{+}) + \ft12 \Delta_+ (u^{+})} \right)
{\rm e}^{\ft12 \Delta_- (u^{+})}
Q^{(4)} (u) = 0
\, . \nonumber
\ea
While using antisymmetric transfer matrices, one finds instead
\ba
\label{BaxterAntiSymQ4}
&&
t_{[2]} (u^+) {\rm e}^{\ft12 \Delta_+ (u^{+})} Q^{(4)} (u + i)
\\
&&\quad
- \,
(x^{-})^L
\left( t_{[1]} (u + i) -  (x^{+})^L {\rm e}^{\ft12 \Delta_- (u^{+}) + \ft12 \Delta_+ (u^{+})} \right)
{\rm e}^{\ft12 \Delta_- (u^{+}) + \Delta_- (u^{-})}
Q^{(4)} (u - i)
\nonumber\\
&&\quad
- \,
\left( t_{[2]} (u^+) - t_{[1]} (u) t_{[1]} (u + i) + (x^{+})^L t_{[1]} (u)
{\rm e}^{\ft12 \Delta_- (u^{+}) + \ft12 \Delta_+ (u^{+})} \right)
{\rm e}^{\ft12 \Delta_- (u^{+})}
Q^{(4)} (u) = 0
\, . \nonumber
\ea
As we see, the Baxter equations for the momentum carrying polynomials completely decouple
from the one built from the nested Bethe roots $u^{(3)}_k$. For the nested Baxter polynomial
$\widehat{Q}^{(3)} (u)$, one finds in analogous manner the TQ-relations in terms of $t_{[a]}$
\be
\label{BaxterAntiSymQ3}
t_{[2]} (u^+) \widehat{Q}^{(3)} (u^+)
=
(x^+)^L {\rm e}^{\ft12 \Delta_+ (u^+) + \ft12 \Delta_- (u^+)}
\left(
(x^+)^L {\rm e}^{\ft12 \Delta_+ (u^+) + \ft12 \Delta_- (u^+)}
-
t_{[1]} (u^{[+ 2]})
\right)
\widehat{Q}^{(3)} (u^-)
\, ,
\ee
and $t^{\{s\}}$ transfer matrices
\ba
\label{BaxterAntiQ3}
\Big(
t^{\{ 1 \}} (u) t^{\{ 1 \}} (u^{[+ 2]})
\!\!\!&-&\!\!\!
t^{\{2\}} (u^+)
\Big)
\widehat{Q}^{(3)} (u^+)
\\
&=&\!\!\!
(x^+)^L {\rm e}^{\ft12 \Delta_+ (u^+) + \ft12 \Delta_- (u^+)}
\left(
(x^+)^L {\rm e}^{\ft12 \Delta_+ (u^+) + \ft12 \Delta_- (u^+)}
-
t^{\{1\}} (u^{[+ 2]})
\right)
\widehat{Q}^{(3)} (u^-)
\, . \nonumber
\ea
It is straightforward to see that both sets of equations, i.e., Eqs.\ \re{BaxterSymQ4},
\re{BaxterAntiSymQ3} and \re{BaxterAntiSymQ4}, \re{BaxterAntiQ3}, are related to each
other by means of the fusion relation for the transfer matrices $t_{[1]} (u^+) t_{[1]}
(u^-) = t_{[2]} (u) + t^{\{2\}} (u)$. Below in Section \ref{sl21diagonalization}, we
perform an explicit diagonalization of these Baxter equations.

\subsection{$sl(2|2)$ sector}

Incorporation of both fermionic nodes of the Kac-Dynkin diagram adjacent to its central node,
with all other being trivial, corresponds to the largest particle-number preserving sector
of the dilatation operator in the maximally supersymmetric gauge theory. Its fermionic and
bosonic excitation content consists of two gauginos $\lambda^1$, $\bar\lambda_4$, and the
complex scalar field $X = \bar{\phi}_{42}$ and the covariant light-cone derivative
$\mathcal{D}_+$, respectively.

To write down the Baxter equations it suffices to use three transfer matrices $t^{\{1\}}(u)$,
$t^{\{2\}}(u)$ and $t^{\{3\}}(u)$ or $t_{[1]}(u)$, $t_{[2]}(u)$ and $t_{[3]}(u)$. However,
contrary to the rank two sector discussed in the previous section, the nesting in the
$sl(2|2)$ case is intrinsic and cannot be eliminated. As a consequence, the eigenvalues for
the momentum carrying Baxter polynomial $Q^{(4)}$ arise from the TQ-relation
\ba
\Bigg(
t_{[1]} (u)
\!\!\!&+&\!\!\!
(x^+)^L {\rm e}^{\ft12 \Delta_+ (u^+) + \ft12 \Delta_- (u^+)}
\frac{\widehat{Q}^{(3)} (u^-)}{\widehat{Q}^{(3)} (u^+)}
+
(x^-)^L {\rm e}^{\ft12 \Delta_- (u^-) + \ft12 \Delta_+ (u^-)}
\frac{\widehat{Q}^{(5)} (u^+)}{\widehat{Q}^{(5)} (u^-)}
\Bigg)
Q^{(4)} (u)
\nonumber\\
&-&\!\!\!
(x^+)^L {\rm e}^{\Delta_+ (u^+)}
\frac{\widehat{Q}^{(3)} (u^-)}{\widehat{Q}^{(3)} (u^+)}
Q^{(4)} (u + i)
-
(x^-)^L {\rm e}^{\Delta_- (u^-)}
\frac{\widehat{Q}^{(5)} (u^+)}{\widehat{Q}^{(5)} (u^-)}
Q^{(4)} (u - i)
= 0
\, . \qquad
\ea
with the dressing factors explicitly depending on the nested Baxter polynomials. In turn,
the nested Baxter functions obey autonomous Baxter equations, which read in terms
of either symmetric
\ba
&&
\left[ t^{\{2\}} (u) t^{\{2\}} (u^{[+ 2]}) - t^{\{1\}} (u^+) t^{\{3\}} (u^+) \right]
{\rm e}^{\ft12 \Delta_+ (u^{+})}
\widehat{Q}^{(3)} (u + i)
\\
&&
+ \,
x^L (x^{[+2]})^L t^{\{1\}} (u^+) t^{\{1\}} (u^{[+ 3]})
{\rm e}^{\ft12 \Delta_+ (u) + \ft12 \Delta_+ (u^{[+ 2]}) + \ft12 \Delta_- (u) + \ft12 \Delta_- (u^{[+ 2]})}
\widehat{Q}^{(3)} (u - i)
\nonumber\\
&&
- \,
(x^{[+2]})^L t^{\{1\}} (u^{[+ 3]}) \left[ t^{\{2\}} (u) - t^{\{1\}} (u^+) t^{\{1\}} (u^-) \right]
{\rm e}^{\ft12 \Delta_- (u^{[+ 2]}) + \ft12 \Delta_+ (u^{[+ 2]})}
\widehat{Q}^{(3)} (u) = 0
\, , \nonumber\\
&&
\left[ t^{\{2\}} (u) t^{\{2\}} (u^{[-2]}) - t^{\{1\}} (u^-) t^{\{3\}} (u^-) \right]
{\rm e}^{\ft12 \Delta_+ (u^{+})}
\widehat{Q}^{(5)} (u - i)
\\
&&
+ \,
x^L (x^{[- 2]})^L t^{\{1\}} (u^-) t^{\{1\}} (u^{[- 3]})
{\rm e}^{\ft12 \Delta_+ (u) + \ft12 \Delta_+ (u^{[- 2]}) + \ft12 \Delta_- (u) + \ft12 \Delta_- (u^{[- 2]})}
\widehat{Q}^{(5)} (u + i)
\nonumber\\
&&
- \,
(x^{[- 2]})^L t^{\{1\}} (u^{[- 3]}) \left[ t^{\{2\}} (u) - t^{\{1\}} (u^+) t^{\{1\}} (u^-) \right]
{\rm e}^{\ft12 \Delta_- (u^{[- 2]}) + \ft12 \Delta_+ (u^{[- 2]})}
\widehat{Q}^{(5)} (u) = 0
\, , \nonumber
\ea
or antisymmetric transfer matrices
\ba
&&
\left[ t_{[2]} (u) t_{[2]} (u^{[+ 2]}) - t_{[1]} (u^+) t_{[3]} (u^+) \right]
\widehat{Q}^{(3)} (u + i)
\\
&&
+ \,
x^L (x^{[+2]})^L t_{[1]} (u^+) t_{[1]} (u^{[+ 3]})
{\rm e}^{\ft12 \Delta_+ (u) + \ft12 \Delta_+ (u^{[+ 2]}) + \ft12 \Delta_- (u) + \ft12 \Delta_- (u^{[+ 2]})}
\widehat{Q}^{(3)} (u - i)
\nonumber\\
&&
+ \,
(x^{[+ 2]})^L t_{[1]} (u^{[+ 3]}) t_{[2]} (u)
{\rm e}^{\ft12 \Delta_- (u^{[+ 2]}) + \ft12 \Delta_+ (u^{[+ 2]})}
\widehat{Q}^{(3)} (u) = 0
\, , \nonumber\\
&&
\left[ t_{[2]} (u) t_{[2]} (u^{[-2]}) - t_{[1]} (u^-) t_{[3]} (u^-) \right]
\widehat{Q}^{(5)} (u - i)
\\
&&
+ \,
x^L (x^{[- 2]})^L t_{[1]} (u^-) t_{[1]} (u^{[- 3]})
{\rm e}^{\ft12 \Delta_+ (u) + \ft12 \Delta_+ (u^{[- 2]}) + \ft12 \Delta_- (u) + \ft12 \Delta_- (u^{[- 2]})}
\widehat{Q}^{(5)} (u + i)
\nonumber\\
&&
+ \,
(x^{[- 2]})^L t_{[1]} (u^{[- 3]}) t_{[2]} (u)
{\rm e}^{\ft12 \Delta_- (u^{[- 2]}) + \ft12 \Delta_+ (u^{[- 2]})}
\widehat{Q}^{(5)} (u) = 0
\, . \nonumber
\ea
As we observe the resulting Baxter equations remain of the second order in finite differences.

\subsection{``Gluonic'' sector}

Let us also address the aligned-helicity gluonic sector which is closed under renormalization
at one loop order \cite{Bel99} (see Refs.\ \cite{RejStaZie07,Bec07} for recent discussion
of related gauge sectors). The space of states of the dilatation operator in this case is
spanned by single-trace Wilson operators built from the twist-one components of the gauge
field strength tensor
\ba
\label{Fproject}
&&
F^{+ \top} \equiv \ft12 \left( F^{+1} - i F^{+2} \right)
= i F_1{}^2 = - i F_{11}
\, , \\
&&
F^{+ \bot} \equiv \ft12 \left( F^{+1} + i F^{+2} \right)
= - i \bar{F}^{\dot{2}}{}_{\dot{1}} = i \bar{F}_{\dot{1}\dot{1}}
\, , \nonumber
\ea
with either all helicity up $F^{+ \top}$ or all down $F^{+ \bot}$ and and arbitrary number $N$
of covariant derivatives $\mathcal{D}_+$,
\be
\label{GluonicOperators}
\tr [ \, \mathcal{D}_+^N (F^{+ \top})^L \, ]
\, , \qquad
\tr [ \, \mathcal{D}_+^N (F^{+ \bot})^L \, ]
\, .
\ee
In the last equalities in Eq.\ \re{Fproject}, we used the spinor decomposition $F^{\mu\nu} =
\ft{1}{\sqrt{2}} \sigma^{\mu\nu}{}_\alpha{}^\beta F_\beta{}^\alpha + \ft{1}{\sqrt{2}}
\bar\sigma^{\mu\nu}{}^{\dot\alpha}{}_{\dot\beta} \bar{F}^{\dot\beta}{}_{\dot\alpha}$ which
clearly identifies helicity content of the involved Lorentz projections.

The embedding of this sector into the $su(2,2|4)$ spin chain requires identification of the
excitation numbers $n_p$ associated with it. The use of general excitation numbers from
Appendix \ref{SuperspaceRealization} yields however states which are superconformal descendants
of \re{GluonicOperators}. Instead, one can use a sequence of particle-hole transformations
starting from the distinguished $su(2,2|4)$ Kac-Dynkin diagram in FFFFBBBB grading (with the
node numbering from left to right) where the above operator corresponds to the excited sixth
node with $\widetilde{n}_6 = N$ and nontrivial Kac-Dynkin labels $\widetilde{w}_5 = 2$ and
$\widetilde{w}_6 = - 3$ on its fifth and and sixth nodes, respectively. Dualizing the one-loop
distinguished Kac-Dynkin diagram to the one associated with the long-range spin chain used in
the present paper, see Fig.\ \ref{Dynkin}, one finds the following nonvanishing excitation
numbers, $n_4 = 2L + N - 2$, $n_5 = 2 L - 4$ and $n_7 = L - 2$. Of course, beyond leading order
of perturbation theory, the gluonic operators get mixed with other Wilson operators of different
field content and possessing identical quantum numbers.

The construction of the Baxter equations follows the steps outlines in the previous two
sections: one writes down the lowest-dimensional transfer matrices and solves them, if possible,
with respect to the nested Baxter polynomials. Eliminating the nested Baxter functions from
the antisymmetric transfer matrices with lowest three dimensions in the auxiliary space, we find
the functional relations for the momentum carrying Baxter polynomials $Q^{(4)} (u)$ which takes
the form a second-order finite-difference equation
\be
d^a_+ (u) Q^{(4)} (u + i) + d^a_- (u) Q^{(4)} (u - i) + d^a_0 (u) Q^{(4)} (u)
=
0
\, ,
\ee
where the dressing factors contain residual dependence on the auxiliary polynomial $Q^{(5)} (u)$,
\ba
d^a_+ (u)
\!\!\!&=&\!\!\!
(x^+)^L {\rm e}^{\ft12 \Delta_+ (u^-) + \Delta_+ (u^+)}
\left\{
(x^{[- 3]})^L
t_{[1]} (u^{[- 2]}) t_{[1]} (u^{[-4]})
{\rm e}^{\ft12 \Delta_- (u^{[- 3]}) + \ft12 \Delta_+ (u^{[- 3]}) + \ft12 \sigma^{(5)} (u^-)}
\right.
\nonumber\\
&&\qquad\qquad\qquad\qquad\qquad
- \,
t_{[2]} (u^{[- 3]})
\left[
(x^{[- 3]})^L
{\rm e}^{\ft12 \Delta_- (u^{[- 3]}) + \ft12 \Delta_+ (u^{[- 3]}) + \ft12 \sigma^{(5)} (u^-)}
\right.
\nonumber\\
&&\qquad\qquad\qquad\qquad\qquad\qquad\qquad\quad
\left.\left.
-
(x^-)^L
{\rm e}^{\ft12 \Delta_- (u^-) + \ft12 \Delta_+ (u^-) + \ft12 \sigma^{(5)} (u^{[- 3]})}
\right]
\right\}
\, , \nonumber\\
d^a_- (u)
\!\!\!&=&\!\!\!
\left[ t_{[2]} (u^-) t_{[2]} (u^{[-3]}) - t_{[1]} (u^{[-2]}) t_{[3]} (u^{[-2]}) \right]
{\rm e}^{\ft12 \Delta_- (u^-) + \ft12 \sigma^{(5)} (u^{[-3]})}
\, , \\
d^a_0 (u)
\!\!\!&=&\!\!\!
{\rm e}^{\ft12 \Delta_+ (u^-)}
\left\{
(x^{[- 3]})^L t_{[1]} (u^{[- 4]})
\left[
t_{[2]} (u^-)
-
t_{[1]} (u) t_{[1]} (u^{[- 2]})
\right]
{\rm e}^{\ft12 \Delta_- (u^{[- 3]}) + \ft12 \Delta_+ (u^{[- 3]}) + \ft12 \sigma^{(5)} (u^-)}
\right.
\nonumber\\
&&
\qquad\qquad\quad
- \left[
t_{[3]} (u^{[- 2]})
-
t_{[1]} (u) t_{[2]} (u^{[- 3]})
\right]
\left[
(x^{[- 3]})^L
{\rm e}^{\ft12 \Delta_- (u^{[- 3]}) + \ft12 \Delta_+ (u^{[- 3]}) + \ft12 \sigma^{(5)} (u^-)}
\right.
\nonumber\\
&&\qquad\qquad\qquad\qquad\qquad\qquad\qquad\qquad\qquad
\left.\left.
-
(x^-)^L
{\rm e}^{\ft12 \Delta_- (u^-) + \ft12 \Delta_+ (u^-) + \ft12 \sigma^{(5)} (u^{[- 3]})}
\right]
\right\}
\, . \nonumber
\ea
Therefore, this have to be supplemented by TQ-relations for the nested Baxter functions
which read
\ba
&&
(x^+)^L {\rm e}^{\Delta_+ (u^+)} d^a_- (u) Q^{(5)} (u^-)
-
(x^-)^L {\rm e}^{\Delta_- (u^-)} d^a_+ (u) Q^{(5)} (u^+)
=
0
\, , \\
&&
(x^-)^L {\rm e}^{\Delta_- (u^-) + \ft12 \sigma^{(5)} (u^+)} d^a_+ (u) Q^{(6)} (u + i)
+
(x^+)^L {\rm e}^{\Delta_+ (u^+) + \ft12 \sigma^{(5)} (u^-)} d^a_- (u) Q^{(6)} (u - i)
\nonumber\\
&&\qquad
+
\,
{\rm e}^{\ft12 \Delta_- (u^-) - \ft12 \Delta_+ (u^-) + \ft12 \sigma^{(5)} (u^-)}
\left[ (x^+)^L {\rm e}^{\Delta_+ (u^+)} d_0^a (u) + t_{[1]} (u) d_+^a (u) \right]
Q^{(6)} (u)
=
0
\, , \nonumber
\ea
Analogously, one finds the Baxter equations for the momentum carrying Baxter polynomials
$Q^{(4)} (u)$ in terms of symmetric transfer matrices
\be
d^s_+ (u) Q^{(4)} (u + i) + d^s_- (u) Q^{(4)} (u - i) + d^s_0 (u) Q^{(4)} (u)
=
0
\, ,
\ee
while the ones for the nested Baxter functions are
\ba
&&
(x^+)^L {\rm e}^{\Delta_+ (u^+)} d^s_- (u) Q^{(5)} (u^-)
-
(x^-)^L {\rm e}^{\Delta_- (u^-)} d^s_+ (u) Q^{(5)} (u^+)
=
0
\, , \\
&&
(x^-)^L {\rm e}^{\Delta_- (u^-) + \ft12 \sigma^{(5)} (u^+)} d^s_+ (u) Q^{(6)} (u + i)
+
(x^+)^L {\rm e}^{\Delta_+ (u^+) + \ft12 \sigma^{(5)} (u^-)} d^s_- (u) Q^{(6)} (u - i)
\nonumber\\
&&\qquad
+
\, {\rm e}^{\ft12 \Delta_- (u^-) - \ft12 \Delta_+ (u^-) + \ft12 \sigma^{(5)} (u^-)}
\left[
(x^+)^L {\rm e}^{\Delta_+ (u^+)} d^s_0 (u)
+
t^{\{1\}} (u) d_+^s (u)
\right]
Q^{(6)} (u)
=
0
\, , \nonumber
\ea
with
\ba
d^s_+ (u)
\!\!\!&=&\!\!\!
(x^+)^L {\rm e}^{\ft12 \Delta_+ (u^-) + \Delta_+ (u^+)}
\left\{
(x^-)^L
t^{\{1\}} (u^{[- 2]}) t^{\{1\}} (u^{[-4]})
{\rm e}^{\ft12 \Delta_- (u^-) + \ft12 \Delta_+ (u^-) + \ft12 \sigma^{(5)} (u^{[- 3]})}
\right.
\nonumber\\
&&\qquad\qquad\qquad\qquad\qquad
+ \,
t^{\{2\}} (u^{[- 3]})
\left[
(x^{[- 3]})^L
{\rm e}^{\ft12 \Delta_- (u^{[- 3]}) + \ft12 \Delta_+ (u^{[- 3]}) + \ft12 \sigma^{(5)} (u^-)}
\right.
\nonumber\\
&&\qquad\qquad\qquad\qquad\qquad\qquad\qquad\quad
\left.\left.
-
(x^-)^L
{\rm e}^{\ft12 \Delta_- (u^-) + \ft12 \Delta_+ (u^-) + \ft12 \sigma^{(5)} (u^{[- 3]})}
\right]
\right\}
\, , \nonumber\\
d^s_- (u)
\!\!\!&=&\!\!\!
\left[ t^{\{2\}} (u^-) t^{\{2\}} (u^{[-3]}) - t^{\{1\}} (u^{[-2]}) t^{\{3\}} (u^{[-2]}) \right]
{\rm e}^{\ft12 \Delta_- (u^-) + \ft12 \sigma^{(5)} (u^{[-3]})}
\, , \\
d^s_0 (u)
\!\!\!&=&\!\!\!
- {\rm e}^{\ft12 \Delta_+ (u^-)}
\left\{
(x^-)^L t^{\{1\}} (u^{[- 4]}) t^{\{2\}} (u^-)
{\rm e}^{\ft12 \Delta_- (u^-) + \ft12 \Delta_+ (u^-) + \ft12 \sigma^{(5)} (u^{[- 3]})}
\right.
\nonumber\\
&&
\qquad\qquad\quad
+
t^{\{3\}} (u^{[- 2]})
\left[
(x^{[- 3]})^L
{\rm e}^{\ft12 \Delta_- (u^{[- 3]}) + \ft12 \Delta_+ (u^{[- 3]}) + \ft12 \sigma^{(5)} (u^-)}
\right.
\nonumber\\
&&\qquad\qquad\qquad\qquad\qquad\qquad\qquad\qquad\qquad
\left.\left.
-
(x^-)^L
{\rm e}^{\ft12 \Delta_- (u^-) + \ft12 \Delta_+ (u^-) + \ft12 \sigma^{(5)} (u^{[- 3]})}
\right]
\right\}
\, . \nonumber
\ea
As anticipated, for vanishing coupling constant, $g = 0$, the nesting disappears indicating
that the aligned-helicity gluon operators form a closed subsector of the dilatation operator
\cite{Bel99}.

\section{Solution of fusion hierarchies}
\label{sl21diagonalization}

Finally, let us address the solution of the fusion hierarchy for the rank-two $sl(2|1)$
sector, i.e., finding explicit eigenvalues for all transfer matrices for a given state
of the spin chain.

The infinite tower of transfer matrices $t_{[a]} (u)$ and $t^{\{s\}} (u)$ can be expressed
in terms of just two low-dimensional antisymmetric transfer matrices $t_{[1]} (u)$ and
$t_{[2]} (u)$. Namely, using the analogue of the duality relation \re{DualityRelation}
for the $sl(2|1)$ sector and solving it with respect to $t_{[3]} (u)$ as in Eq.\
\re{SolutionDualityRel}, one finds
\ba
\label{ta3}
t_{[3]} (u)
\!\!\!&=&\!\!\!
\frac{t_{[2]} (u^-) t_{[2]} (u^+)}{t_{[1]} (u) - (x^-)^L
{\rm e}^{\ft12 \Delta_+ (u^-) + \ft12 \Delta_- (u^-)}}
\\
&=&\!\!\!
-
(x^{[- 1]})^L
{\rm e}^{
\ft12 \Delta_+ (u^{[- 1]}) + \ft12 \Delta_- (u^{[- 1]})
}
t_{[2]} (u^+)
\frac{
\widehat{Q}^{(3)} (u^{[- 3]})
}{
\widehat{Q}^{(3)} (u^{[- 1]})
}
\, . \nonumber
\ea
All other transfer matrices with antisymmetric auxiliary space can be found iteratively
\be
\label{tan}
t_{[a + 1]} (u)
=
- (x^{[- a + 1]})^L
{\rm e}^{
\ft12 \Delta_+ (u^{[- a + 1]}) + \ft12 \Delta_- (u^{[- a + 1]})
}
t_{[a]} (u^+)
\frac{
\widehat{Q}^{(3)} (u^{[- a - 1]})
}{
\widehat{Q}^{(3)} (u^{[- a + 1]})
}
\, ,
\ee
for $a > 3$. While all transfer matrices with symmetric representations in the auxiliary
space follow from the Bazhanov-Reshetikhin formula \re{BazhanovReshetikhin}.

Therefore, the problem is reduced to the solution of functional relations \re{BaxterAntiSymQ4}
and \re{BaxterAntiSymQ3} for Baxter polynomials $Q^{(4)} (u)$ and $\widehat{Q}^{(3)} (u)$ and
transfer matrices $t_{[1]} (u)$ and $t_{[2]} (u)$. At leading order of perturbations theory,
i.e., for $g = 0$, Eq.\ \re{BaxterAntiSymQ4} can be solved immediately along the same lines
as the $sl(2)$ Baxter equation analyzed in Appendix \ref{SolutBaxter}. However, beyond one
loop its successful solution requires proper identification of the renormalized spectral
parameter which serves as an expansion variable for polynomial fused transfer matrices. The
reason for this is that the perturbative series for the transfer matrix $t_{[a]}$ acquires
nonpolynomial terms in the bare spectral parameter $u$. This happens before the wrapping order
$\mathcal{O} (g^{a L})$ sets in, when the asymptotic Baxter equation is expected to work and
this is ensured by the right choice of the expansion variable. A guidance into the acceptable
choice can be gained by comparing Baxter equations in terms of fused transfer matrices with
the ones written by means of conjugate fundamental transfer matrices introduced in Ref.\
\cite{Bel07a}. Namely, in terms of the transfer matrix $t_{[1]}$ and its conjugate $\bar{t}_{[1]}$,
one finds instead of Eq.\ \re{BaxterAntiSymQ3}
\be
\label{BaxterConjugateQ3}
\left(
(x^-)^L {\rm e}^{\ft12 \Delta_+ (u^-) + \ft12 \Delta_- (u^-)}
-
t_{[1]} (u)
\right)
\widehat{Q}^{(3)} (u^+)
=
\left(
(x^+)^L {\rm e}^{\ft12 \Delta_+ (u^+) + \ft12 \Delta_- (u^+)}
-
\bar{t}_{[1]} (u)
\right)
\widehat{Q}^{(3)} (u^-)
\, .
\ee
Eliminating $\widehat{Q}^{(3)}$ from Eqs.\ \re{BaxterAntiSymQ3} and \re{BaxterConjugateQ3},
one deduces
\be
\label{t2int1t1bar}
t_{[2]} (u)
=
x^L {\rm e}^{\ft12 \Delta_+ (u) + \ft12 \Delta_- (u)}
\left(\!
(x^{[- 2]})^L {\rm e}^{\ft12 \Delta_+ (u^{[- 2]}) + \ft12 \Delta_- (u^{[- 2]})}
-
t_{[1]} (u^-)
\!\right)
\frac{
x^L {\rm e}^{\ft12 \Delta_+ (u) + \ft12 \Delta_- (u)}
-
t_{[1]} (u^+)
}{
x^L {\rm e}^{\ft12 \Delta_+ (u) + \ft12 \Delta_- (u)}
-
\bar{t}_{[1]} (u^-)
}
,
\ee
which suggests that the transfer matrix $t_{[2]} (x)$ can be factorized into a product of two
functions, one depending on $x$ and another on $x^{[- 2]}$. For states with real charges
$\mathfrak{Q}^{[1]}_k$ entering $t_{[1]} (u)$, $\Im {\rm m} \,[ \mathfrak{Q}^{[1]}_k ] = 0$,
it further simplifies to
\baa
t_{[2]} (u)
=
x^L {\rm e}^{\ft12 \Delta_+ (u) + \ft12 \Delta_- (u)}
\left(
(x^{[- 2]})^L {\rm e}^{\ft12 \Delta_+ (u^{[- 2]}) + \ft12 \Delta_- (u^{[- 2]})}
-
t_{[1]} (u^-)
\right)
\, .
\eaa
Taking into account these considerations, the transfer matrices admit the following
expansions in the renormalized spectral parameters
\ba
\label{ta1}
t_{[1]} (u)
\!\!\!&=&\!\!\!
\sum_{k \geq 0} \mathfrak{Q}^{[1]}_k (g) \, (x^-)^{L - k}
\, , \\
\label{ta2}
t_{[2]} (u)
\!\!\!&=&\!\!\!
\sum_{j \geq 0} x^{L - j} \, \mathfrak{R}^{[2]}_j (g) \,
\sum_{k \geq 2} \mathfrak{Q}^{[2]}_k (g) \, (x^{[- 2]})^{L - k}
\, ,
\ea
where the tower of $\mathfrak{R}^{[2]}_j-$charges encodes the expansion of the
exponential dressing factors in Eq.\ \re{t2int1t1bar}. The first few charges in both
transfer matrices \re{ta1} and \re{ta2} can be found in a closed form in terms of
the Baxter polynomial $Q^{(4)} (u)$ by studying the asymptotics of $t_{[1]} (u)$
and $t_{[2]} (u)$ via the fusion relations \re{TransferAntiSym}. This requires the
large$-u$ behavior of the dressing factors $\Delta_\pm (u)$. Using Eqs.\ \re{Sigmap}
and \re{ThetaWeak}, one finds that the first few terms in the $1/u-$expansion yields
\be
\label{ResduesDressing}
\Delta_\pm (u)
=
\frac{\Delta^{(1)}_\pm (g)}{u}
+
\frac{\Delta^{(2)}_\pm (g)}{u^2}
+
\mathcal{O} (u^{- 3})
\, ,
\ee
where the residues are $(\alpha = 1, 2)$
\be
\Delta^{(\alpha)}_\pm =
\int_{- 1}^1 \frac{d t}{\pi}
\sqrt{1 - t^2}
\left\{
w^{(\alpha)} (g t, g)
\left( \ln Q^{(4)} ( \pm \ft{i}{2} - g t ) \right)^\prime
-
\vartheta^{(\alpha)} (g t, g)
\left(
\ln
\frac{
Q^{(4)} ( + \ft{i}{2} - g t )
}{
Q^{(4)} ( - \ft{i}{2} - g t )
}
\right)^\prime
\right\}
\, ,
\ee
and the explicit form of the functions entering the integrand is
\ba
&&
w^{(1)} (t, g) = - g^2
\, , \qquad
\vartheta^{(1)} (t, g) = 32 i t \frac{\mathcal{Z}_{2,1} (g)}{g^2}
\, , \\
&&
w^{(2)} (t, g) = \ft12 g^2 t
\, , \qquad
\vartheta^{(2)} (t, g) = 8 i \mathcal{Z}_{2,1} (g)
\, , \nonumber
\ea
with \cite{BeiEdeSta06,BenBenKleSca06}
\be
\mathcal{Z}_{2,1} (g) = \left( \frac{g}{2} \right)^3
\int_0^\infty dv \frac{J_1 (gv) J_2 (gv)}{v ({\rm e}^{v} - 1)}
\, .
\ee
From these considerations, one immediately concludes that $\mathfrak{Q}^{[a]}_0 (g) =
\mathfrak{R}^{[a]}_0 (g) = 1$ for both $a = 1, 2$ and the all-order conserved charges
$\mathfrak{Q}^{[1]}_1$ and  $\mathfrak{Q}^{[1]}_2$ of the fundamental transfer matrix
$t_{[1]} (u)$ read
\ba
\mathfrak{Q}^{[1]}_1 (g)
\!\!\!&=&\!\!\!
\ft{1}{2}
\left( \Delta_+^{(1)} (g) + \Delta_-^{(1)} (g) \right)
\, , \\
\mathfrak{Q}^{[1]}_2 (g)
\!\!\!&=&\!\!\!
-
\mathfrak{Q}^{[2]}_2 (g)
+
\ft12 \left( \mathfrak{Q}^{[1]}_2 (g) \right)^2
+
\ft12 \left( \Delta_+^{(2)} (g) + \Delta_-^{(2)} (g) \right)
\, ,
\ea
with the latter charge expressed in terms of $\mathfrak{Q}^{[2]}_k$ introduced below. On
the other hand, for the renormalized fused transfer matrix $T_{[2]} (u) = x^{- L} t_{[2]}
(u)$, one deduces by the same token that the leading asymptotics as $u \to \infty$
\be
T_{[2]} (u) = \mathbb{C}_2 (g) u^{L - 2} + \mathcal{O} (u^{L - 3})
\, ,
\ee
is driven by the value of the renormalized quadratic Casimir of the $sl(2|1)$ algebra
\be
\label{sl21quadraticCasimir}
\mathbb{C}_2 (g)
=
\left( n_4 + \ft12 \gamma (g) \right)
\left( n_4 - n_3 + L - 1 + \ft12 \gamma (g) \right)
\, ,
\ee
where the anomalous dimension \re{AllOrderAD} is expressed in terms of residues of the
dressing factors \re{ResduesDressing}
\be
\gamma (g) = - i
\left( \Delta_+^{(1)} (g) - \Delta_-^{(1)} (g) \right)
\, .
\ee
This charge, as we saw above, defines up to an additive contribution, the subleading term
in the fundamental transfer matrix. With this knowledge at hand, we can understand why the
expression in braces in Eq.\ \re{t2int1t1bar} does not have the induced charge
$\mathfrak{Q}^{[1]}_1$: it cancels between the two terms in the large$-u$ limit, however,
it re-appears in the multiplicative factor in front of the polynomial in $x^{[-2]}$ variable,
see Eq.\ \re{ta2}. So that
\be
\mathfrak{Q}^{[2]}_2 (g)
=
\mathbb{C}_2 (g)
\, , \qquad
\mathfrak{R}^{[2]}_1 (g)
=
\mathfrak{Q}^{[2]}_1 (g)
\, .
\ee

This information suffices to solve the multi-loop Baxter equation to arbitrary\footnote{We
implicitly imply that it holds only below the wrapping order due to the asymptotic character
of Baxter equations.} order of perturbation theory. In Table \ref{ExactSpectra} we give,
as an example, the lowest four orders in the expansion of the anomalous dimensions and
conserved charges
\be
\gamma_{\bit{\scriptstyle\alpha}} (g)
=
\sum_{\ell = 0}
g^{2 \ell + 2} \gamma_{\ell,\bit{\scriptstyle\alpha}}
\, , \qquad
\mathfrak{Q}^{[1]}_k (g)
=
\sum_{\ell = 0}
g^{2 \ell}
\mathfrak{Q}^{[1]}_{\ell, k}
\, ,
\ee
for the state $\bit{\alpha} \equiv [L, n_3, n_4] = [5, 3, 7]$ of the $sl(2|1)$ long-range
spin chain. Conserved charges associated with other fused transfer matrices can be found
by means of Eqs.\ \re{t2int1t1bar}, \re{ta3} and \re{tan}.

\begin{table}
\renewcommand{\arraystretch}{1.5}
\begin{center}
\begin{tabular}[pos]{||c|c|c|c|c||}
\hline \hline
$k$ & $0$ & $1$ & $2$ & $3$
\\
\hline \hline
$\gamma_{\bit{\scriptstyle\alpha}} (g)$
&
$
\begin{array}{c}
7
\\
\frac{19}{3}
\end{array}
$
&
$
\begin{array}{c}
- \frac{679}{96}
\\
- \frac{5587}{864}
\end{array}
$
&
$
\begin{array}{c}
\frac{59297}{4608}
\\
\frac{735545}{62208}
\end{array}
$
&
$
\begin{array}{c}
\frac{6355979}{221184} + \frac{175}{96} \zeta (3)
\\
\frac{473456465}{17915904} + \frac{493}{288} \zeta (3)
\end{array}
$
\\
\hline
$
\mathfrak{Q}^{[1]}_1 (g)
$
&
$
\begin{array}{c}
0
\\
0
\end{array}
$
&
$
\begin{array}{c}
\ft{\sqrt{21}}{12}
\\
\ft{\sqrt{7}}{12}
\end{array}
$
&
$
\begin{array}{c}
- \ft{83 \sqrt{21}}{1152}
\\
- \ft{1025 \sqrt{7}}{12096}
\end{array}
$
&
$
\begin{array}{c}
\ft{27703 \sqrt{21}}{221184} + \ft{\sqrt{21}}{48} \zeta(3)
\\
\ft{3811897 \sqrt{7}}{24385536} - \ft{\sqrt{7}}{144} \zeta(3)
\end{array}
$
\\
\hline
$
\mathfrak{Q}^{[1]}_2 (g)
$
&
$
\begin{array}{c}
- 56
\\
- 56
\end{array}
$
&
$
\begin{array}{c}
- \ft{105}{2}
\\
- \ft{95}{2}
\end{array}
$
&
$
\begin{array}{c}
\ft{1295}{32}
\\
\ft{229}{6}
\end{array}
$
&
$
\begin{array}{c}
- \ft{36267}{512} - \ft{7}{8} \zeta(3)
\\
- \ft{465511}{6912} - \ft{19}{24} \zeta(3)
\end{array}
$
\\
\hline
$
\mathfrak{Q}^{[1]}_3 (g)
$
&
$
\begin{array}{c}
-
8 \sqrt{21}
\\
- 16 \sqrt{7}
\end{array}
$
&
$
\begin{array}{c}
- \ft{79 \sqrt{21}}{6}
\\
- \ft{845 \sqrt{7}}{42}
\end{array}
$
&
$
\begin{array}{c}
\ft{12887 \sqrt{21}}{2304}
\\
\ft{268805 \sqrt{7}}{21168}
\end{array}
$
&
$
\begin{array}{c}
- \ft{556337 \sqrt{21}}{55296} - \ft{35 \sqrt{21}}{24} \zeta (3)
\\
- \ft{1862658941 \sqrt{7}}{85349376} - \ft{25 \sqrt{7}}{72} \zeta (3)
\end{array}
$
\\
\hline
$
\mathfrak{Q}^{[1]}_4 (g)
$
&
$
\begin{array}{c}
70
\\
38
\end{array}
$
&
$
\begin{array}{c}
\ft{329}{2}
\\
\ft{547}{6}
\end{array}
$
&
$
\begin{array}{c}
- \ft{2933}{192}
\\
- \ft{11401}{1728}
\end{array}
$
&
$
\begin{array}{c}
\ft{528031}{9216} + \ft{63}{4} \zeta(3)
\\
\ft{3761615}{124416} + \ft{433}{36} \zeta(3)
\end{array}
$
\\
\hline
$
\mathfrak{Q}^{[1]}_5 (g)
$
&
$
\begin{array}{c}
8 \sqrt{21}
\\
0
\end{array}
$
&
$
\begin{array}{c}
\ft{155 \sqrt{21}}{6}
\\
0
\end{array}
$
&
$
\begin{array}{c}
\ft{19865 \sqrt{21}}{2304}
\\
0
\end{array}
$
&
$
\begin{array}{c}
\ft{61963 \sqrt{21}}{55296} + \ft{7 \sqrt{21}}{2} \zeta (3)
\\
0
\end{array}
$
\\
\hline \hline
\end{tabular}
\end{center}
\caption{\label{ExactSpectra} Eigenvalues of the $\bit{\alpha} = [5,3,7]$ states up to
four-loop order in $\mathcal{N} = 4$ super-Yang-Mills theory.}
\end{table}

\section{Conclusions}

In this work, we have suggested the hierarchy of fused transfer matrices for the
putative long-range magnet of the maximally supersymmetric gauge theory. These allow
one to construct a finite set of Baxter equations for eigenvalues of Baxter polynomials
for momentum carrying and auxiliary Bethe roots. We have demonstrated the formalism
using low-rank subsectors of the dilatation operator.

The T-system of transfer matrices formulated in Section \ref{Tsystem} is a starting
point for the derivation of the Y-systems \cite{BazRes89,KunOhtSuz95,KunNakSuz93} which,
in turn, yield equations of the Thermodynamic Bethe Ansatz \cite{KluPea92,JutKluSuz98}.
The latter, implemented for the mirror image of the long-range spin chain, incorporate
finite-size effects \cite{Jan05,AruFro07} with the lowest order correction reproducing
the L\"uscher formula (see, e.g., Ref.\ \cite{BalHeg03}), which was recently reincarnated
within the gauge/string duality in Refs.\ \cite{Jan05} to account for wrapping effects
evading the framework of the asymptotic Bethe Ansatz. Therefore, one anticipates that
the framework outlined here may be used to consistently incorporate this kind of
finite-size effects.

\vspace{0.5cm}

\noindent This work was supported by the U.S. National Science Foundation under grant
no. PHY-0456520. We are indebted to Romuald Janik, Gregory Korchemsky, Vladimir Korepin,
Rafael Nepomechie, Adam Rej, Matthias Staudacher, Zengo Tsuboi and Stefan Zieme for
useful discussions and correspondence at different stages of this work.

\appendix

\section{Superconformal algebra}
\label{SuperspaceRealization}

Let us present the Serre-Chevalley basis associated with the Kac-Dynkin diagram in Fig.\
\ref{Dynkin}. As a corollary of our considerations we will identify the excitation numbers
defining Baxter polynomials with quantum numbers of gauge invariant Wilson operators that
they correspond to. We start with the $su(2,2|4)$ superconformal algebra in the notation of
Ref.\ \cite{Bel07b} and use spinor projections
\be
\label{SuperConformalGenerators}
\begin{array}{lllll}
\mathbb{L}_\alpha{}^\beta
=
- \ft14 \mathfrak{M}_{\mu\nu} \sigma^{\mu\nu}{}_\alpha{}^\beta
\, ,
&
\bar{\mathbb{L}}^{\dot\alpha}{}_{\dot\beta}
=
\ft14 \mathfrak{M}^{\mu\nu}
\bar\sigma_{\mu\nu}{}^{\dot\alpha}{}_{\dot\beta}
\, ,
&
\mathbb{P}_{\alpha\dot\beta}
= \ft{i}{2}
\mathfrak{P}_\mu
\bar\sigma^\mu{}_{\alpha\dot\beta}
\, ,
&
\mathbb{K}^{\dot\alpha\beta}
=
\ft{i}{2}
\mathfrak{K}^\mu
\sigma_\mu{}^{\dot\alpha\beta}
\, ,
&
\mathbb{D} = i \mathfrak{D}
\, , \\ [1mm]
\mathbb{Q}_{\alpha A}
=
\ft{i}{2} \mathfrak{Q}_{\alpha A}
\, ,
&
\bar{\mathbb{Q}}_{\dot\beta}^B
=
\ft12 \bar{\mathfrak{Q}}_{\dot\beta}^B
\, ,
&
\mathbb{S}_\alpha^A
=
\ft{i}{2} \mathfrak{S}_\alpha^A
\, ,
&
\bar{\mathbb{S}}^{\dot\beta}_A
=
\ft12
\bar{\mathfrak{S}}^{\dot\beta}_A
\, ,
&
\mathbb{T}_A{}^B = \mathfrak{T}_A{}^B
\, . \!\!\!\!
\end{array}
\ee
The chirality charge $\mathbb{R}$ is merely identified $\mathbb{R} = \mathfrak{R}$. The
inverse transformations for the bosonic generators of the $so(4,2)$ conformal algebra are
\ba
\mathfrak{M}^{\mu\nu}
=
- \ft12 \tr \left[  \sigma^{\mu\nu} \, \mathbb{L} \right]
+ \ft12 \tr \left[ \bar{\sigma}^{\mu\nu} \, \bar{\mathbb{L}}  \right]
\, , \qquad
i \mathfrak{P}^\mu = \tr \left[ \sigma^\mu \, \mathbb{P} \right]
\, , \qquad
i \mathfrak{K}^\mu = \tr \left[ \bar{\sigma}^\mu \, \mathbb{K} \right]
\, .
\ea
The generators \re{SuperConformalGenerators} can be organized in a supermatrix
\be
\mathbb{E}^{\mathcal{A}}{}_{\mathcal{B}}
=
\left(
\begin{array}{ccc}
- \bar{\mathbb{L}}^{\dot\alpha}{}_{\dot\beta}
-
\ft12 \delta^{\dot\alpha}{}_{\dot\beta}
\left(
\mathbb{D} + \ft12 \mathbb{R}
\right)
\
&
- \bar{\mathbb{S}}^{\dot\alpha}_B
\
&
- \mathbb{K}^{\dot\alpha\beta}
\
\\ [3mm]
\bar{\mathbb{Q}}_{\dot\beta}^A
\
&
- \mathbb{T}_B{}^A + \ft14 \delta_B{}^A \mathbb{R}
\
&
\mathbb{S}^{\beta A}
\
\\ [3mm]
\mathbb{P}_{\alpha\dot\beta}
\
&
\mathbb{Q}_{\alpha B}
\
&
\mathbb{L}_\alpha{}^\beta
+
\ft12 \delta_\alpha{}^\beta
\left(
\mathbb{D} - \ft12 \mathbb{R}
\right)
\end{array}
\right)
\, ,
\ee
with indices running over eight values $\mathcal{A} = (\dot\alpha, A, \alpha)$. They obey
the graded commutation relations
\ba
\label{GradedCommutator}
{}[
\mathbb{E}^{\mathcal{A}}{}_{\mathcal{B}} , \mathbb{E}^{\mathcal{C}}{}_{\mathcal{D}}
\}
\!\!\!&\equiv&\!\!\!
\mathbb{E}^{\mathcal{A}}{}_{\mathcal{B}} \mathbb{E}^{\mathcal{C}}{}_{\mathcal{D}}
-
(-1)^{(\bar{\mathcal{A}} + \bar{\mathcal{B}})(\bar{\mathcal{C}} + \bar{\mathcal{D}})}
\mathbb{E}^{\mathcal{C}}{}_{\mathcal{D}} \mathbb{E}^{\mathcal{A}}{}_{\mathcal{B}}
\nonumber\\
&=&\!\!\!
\delta^{\mathcal{C}}{}_{\mathcal{B}} \mathbb{E}^{\mathcal{A}}{}_{\mathcal{D}}
-
(-1)^{(\bar{\mathcal{A}} + \bar{\mathcal{B}})(\bar{\mathcal{C}} + \bar{\mathcal{D}})}
\delta^{\mathcal{A}}{}_{\mathcal{D}} \mathbb{E}^{\mathcal{C}}{}_{\mathcal{B}}
\, .
\ea

The root system of the $su(2,2|4)$ algebra is expressed in terms of weights which form a
basis of the dual Cartan subalgebra. We enumerate them as follows
\be
\bit{v}_\alpha = (\bit{\varepsilon}_3, \bit{\varepsilon}_4
| \bit{\delta}_1, \bit{\delta}_2, \bit{\delta}_3, \bit{\delta}_4|
\bit{\varepsilon}_1, \bit{\varepsilon}_2)
\, ,
\ee
and endow them with inner products
\be
( \bit{\varepsilon}_i | \bit{\varepsilon}_j ) = \delta_{ij}
\, , \qquad
( \bit{\delta}_i | \bit{\delta}_j ) = - \delta_{ij}
\, , \qquad
( \bit{\varepsilon}_i | \bit{\delta}_j ) = 0
\, .
\ee
The simple root system corresponding to the Kac-Dynkin diagram in Fig.\ \ref{Dynkin} reads
\ba
&&
\bit{\alpha}_1 = \bit{\varepsilon}_2 - \bit{\delta}_1
\, , \quad
\bit{\alpha}_2 = \bit{\delta}_1 - \bit{\delta}_2
\, , \quad
\bit{\alpha}_3 = \bit{\delta}_2 - \bit{\varepsilon}_1
\, , \quad
\bit{\alpha}_4 = \bit{\varepsilon}_1 - \bit{\varepsilon}_3
\, , \nonumber\\
&&
\qquad
\bit{\alpha}_5 = \bit{\varepsilon}_3 - \bit{\delta}_3
\, , \qquad
\bit{\alpha}_6 = \bit{\delta}_3 - \bit{\delta}_4
\, , \qquad
\bit{\alpha}_7 = \bit{\delta}_4 - \bit{\varepsilon}_4
\, ,
\ea
and the resulting Cartan matrix $A_{pq} = (\bit{\alpha}_p | \bit{\alpha}_q)$
\be
A =
\left(
\begin{array}{rrrrrrr}
  &  1 &    &    &    &    &   \\
1 & -2 & 1  &    &    &    &   \\
  &  1 &    & -1 &    &    &   \\
  &    & -1 &  2 & -1 &    &   \\
  &    &    & -1 &    &  1 &   \\
  &    &    &    &  1 & -2 & \phantom{-}1 \\
  &    &    &    &    &  1 &
\end{array}
\right)
\, ,
\ee
defines the Serre-Chevalley basis of the superconformal algebra
\be
\begin{array}{lll}
h_1 = \mathbb{E}^8{}_8 + \mathbb{E}^3{}_3 = - \mathbb{L}_1{}^1 + \ft12 \mathbb{D} - \mathbb{T}_1{}^1
\, , \quad &
e_1^+ = \mathbb{E}^8{}_3 = \mathbb{Q}_{21}
\, , \quad &
e_1^- = \mathbb{E}^3{}_8 = \mathbb{S}^{21}
\, ,
\\ [1mm]
h_2 = - \mathbb{E}^3{}_3 + \mathbb{E}^4{}_4 = \mathbb{T}_1{}^1 - \mathbb{T}_2{}^2
\, , \quad &
e_2^+ = \mathbb{E}^3{}_4 = - \mathbb{T}_2{}^1
\, , \quad &
e_2^- = \mathbb{E}^4{}_3 = - \mathbb{T}_1{}^2
\, ,
\\ [1mm]
h_3 = - \mathbb{E}^4{}_4 - \mathbb{E}^7{}_7 = - \mathbb{L}_1{}^1 - \ft12 \mathbb{D} + \mathbb{T}_2{}^2
\, , \quad &
e_3^+ = \mathbb{E}^4{}_7 = \mathbb{S}^{12}
\, , \quad &
e_3^- = \mathbb{E}^7{}_4 = \mathbb{Q}^{12}
\, ,
\\ [1mm]
h_4 = \mathbb{E}^7{}_7 - \mathbb{E}^1{}_1 = \mathbb{L}_1{}^1 + \bar{\mathbb{L}}^{\dot{1}}{}_{\dot{1}} + \mathbb{D}
\, , \quad &
e_4^+ = \mathbb{E}^7{}_1 = \mathbb{P}_{1\dot{1}}
\, , \quad &
e_4^- = \mathbb{E}^1{}_7 = - \mathbb{K}^{\dot{1}1}
\, ,
\\ [1mm]
h_5 = \mathbb{E}^1{}_1 + \mathbb{E}^5{}_5 = - \bar{\mathbb{L}}^{\dot{1}}{}_{\dot{1}} - \ft12 \mathbb{D} - \mathbb{T}_3{}^3
\, , \quad &
e_5^+ = \mathbb{E}^1{}_5 = - \bar{\mathbb{S}}^{\dot{1}}_3
\, , \quad &
e_5^- = \mathbb{E}^5{}_1 = \bar{\mathbb{Q}}^3_{\dot{1}}
\, ,
\\ [1mm]
h_6 = - \mathbb{E}^5{}_5 + \mathbb{E}^6{}_6 = \mathbb{T}_3{}^3 - \mathbb{T}_4{}^4
\, , \quad &
e_6^+ = \mathbb{E}^5{}_6 = - \mathbb{T}_4{}^3
\, , \quad &
e_6^- = \mathbb{E}^6{}_5 = - \mathbb{T}_3{}^4
\, ,
\\ [1mm]
h_7 = \mathbb{E}^2{}_2 - \mathbb{E}^6{}_6 = \bar{\mathbb{L}}^{\dot{1}}{}_{\dot{1}} - \ft12 \mathbb{D} + \mathbb{T}_4{}^4
\, , \quad &
e_7^+ = \mathbb{E}^6{}_2 = \bar{\mathbb{Q}}_{\dot{2}}^4
\, , \quad &
e_7^- = \mathbb{E}^2{}_6 = - \bar{\mathbb{S}}^{\dot{2}}_4
\, ,
\end{array}
\ee
via the conventional commutation relations $[h_p, e^\pm_q] = \pm A_{pq} e^\pm_q$.
Notice that other simple root systems are related to this one by reflections
with respect to the fermionic simple roots $\bit{\alpha}_p$ ($p = 1,3,5,7$) with
the vanishing bilinear form $(\bit{\alpha}|\bit{\alpha}) = 0$. They span the Weyl
supergroup. For Bethe Ansatz equations these correspond to particle-hole transformations.

Making use of graded commutation relations \re{GradedCommutator}, one can cast the
generators \re{SuperConformalGenerators} into the oscillator representation \cite{GunSac82}.
The latter is built from the bosonic $(a^\alpha, a^\dagger_\alpha, b^{\dot\beta},
b^\dagger_{\dot\beta})$ and fermionic $(c_A, c^{\dagger A})$ creation/annihilation
operators and reads
\ba
&&
\mathbb{P}_{\alpha \dot{\beta}} = a_\alpha^\dagger b_{\dot{\beta}}^\dagger
\, , \quad
\mathbb{K}^{\dot{\alpha} \beta} = a^\beta b^{\dot{\alpha}}
\, , \quad
\mathbb{D} = \ft12 a_\gamma^\dagger a^\gamma + \ft12 b_{\dot{\gamma}}^\dagger b^{\dot{\gamma}} + 1
\, , \\
&&
\mathbb{L}_\alpha{}^\beta
=
a_\alpha^\dagger a^\beta
-
\ft12 \delta_\alpha^\beta a_\gamma^\dagger a^\gamma
\, , \qquad
\bar{\mathbb{L}}^{\dot{\alpha}}{}_{\dot{\beta}}
=
b_{\dot{\beta}}^\dagger b^{\dot{\alpha}}
-
\ft12 \delta_{\dot{\alpha}}^{\dot{\beta}} b_{\dot{\gamma}}^\dagger b^{\dot{\gamma}}
\, , \nonumber\\
&&
\mathbb{Q}_{\alpha A} = c_A a_\alpha^\dagger
\, , \quad
\bar{\mathbb{Q}}^A_{\dot{\alpha}} = c^{\dagger A} b_{\dot\alpha}^\dagger
\, , \quad
\mathbb{S}^{\alpha A} = c^{\dagger A} a^\alpha
\, , \quad
\bar{\mathbb{S}}_A^{\dot{\alpha}} = c_A b^{\dot\alpha}
\, , \nonumber\\ [2mm]
&&
\mathbb{T}_A{}^B = c_A c^{\dagger B} - \ft14 \delta_A^B c_D c^{\dagger D}
\, ,
\qquad
\mathbb{R} = b_{\dot{\gamma}}^\dagger b^{\dot{\gamma}} - a_\gamma^\dagger a^\gamma
\, ,
\nonumber
\ea
with the vanishing central charge condition $c^{\dagger D} c_D + a_\gamma^\dagger a^\gamma -
b_{\dot{\gamma}}^\dagger b^{\dot{\gamma}} = 2$. This representation allows one to easily
identify the excitation numbers $n_p$ with eigenvalues of Cartan generators. Introducing a
generic state of the spin chain,
\ba
| \mathcal{O}_{n_1, \dots , n_7} \rangle
\!\!\!&=&\!\!\!
\left( \prod_{k = 1,2,3}^{\curvearrowleft} (e^+_{4 + k})^{n_{4 + k}} (e^+_{4 - k})^{n_{4 - k}} \right)
(e^+_4)^{n_4} | \Omega \rangle
\\
\!\!\!&\sim&\!\!\!
(c_4)^{n_6 - n_7} (c_3)^{n_5 - n_6} (c^{\dagger 1})^{n_2 - n_1} (c^{\dagger 2})^{n_3 - n_2}
(a_1^\dagger)^{n_4 - n_3} (a_2^\dagger)^{n_1} (b_{\dot{1}}^\dagger)^{n_4 - n_5} (b_{\dot{2}}^\dagger)^{n_7}
| \Omega \rangle
\, , \qquad\nonumber
\ea
where we remind that the vacuum state is $| \Omega \rangle = (c^{\dagger 3} c^{\dagger 4})^L |
0 \rangle = \tr [\, Z^L (0) \,]$, one observes that the excitation numbers $n_p$ can be ``counted''
with the Cartan generators by relating them to the number-of-particle operators
\ba
\begin{array}{ll}
a_1^\dagger a^1
=
\ft12 \mathbb{D} + \mathbb{L}_1{}^1 - \ft14 \mathbb{R} - \ft12
\, ,
&
\qquad
a_2^\dagger a^2
=
\ft12 \mathbb{D} - \mathbb{L}_1{}^1 - \ft14 \mathbb{R} - \ft12
\, , \\ [2mm]
b_{\dot{1}}^\dagger b^{\dot{1}}
=
\ft12 \mathbb{D} + \bar{\mathbb{L}}^{\dot{1}}{}_{\dot{1}} + \ft14 \mathbb{R} - \ft12
\, ,
&
\qquad
b_{\dot{2}}^\dagger b^{\dot{2}}
=
\ft12 \mathbb{D} - \bar{\mathbb{L}}^{\dot{1}}{}_{\dot{1}} + \ft14 \mathbb{R} - \ft12
\, , \\ [2mm]
c^{\dagger 1} c_1
=
- \mathbb{T}_1{}^1 + \ft14 \mathbb{R} + \ft12
\, ,
&
\qquad
c^{\dagger 2} c_2
=
- \mathbb{T}_2{}^2 + \ft14 \mathbb{R} + \ft12
\, , \\ [2mm]
c_3 c^{\dagger 3}
=
\mathbb{T}_3{}^3 - \ft14 \mathbb{R} + \ft12
\, ,
&
\qquad
c_4 c^{\dagger 4}
=
\mathbb{T}_4{}^4 - \ft14 \mathbb{R} + \ft12
\, ,
\end{array}
\ea
where $\sum_{i = 1}^4 \mathbb{T}_i{}^i = 0$. This yields the following identities
\ba
n_1
\!\!\!&=&\!\!\!
\ft12 d - s - \ft14 r - \ft12 L
\, , \nonumber\\
n_2
\!\!\!&=&\!\!\!
\ft12 d - s - t_1
\, , \nonumber\\
n_3
\!\!\!&=&\!\!\!
\ft12 d - s + \ft14 r - t_1 - t_2 + \ft12 L
\, , \nonumber\\
n_4
\!\!\!&=&\!\!\!
d - t_1 - t_2
\, , \\
n_5
\!\!\!&=&\!\!\!
\ft12 d - \bar{s} - \ft14 r + t_3 + t_4 + \ft12 L
\, , \nonumber\\
n_6
\!\!\!&=&\!\!\!
\ft12 d - \bar{s} +  t_4
\, , \nonumber\\
n_7
\!\!\!&=&\!\!\!
\ft12 d - \bar{s} + \ft14 r - \ft12 L
\, , \nonumber
\ea
where $d$, $s$, $\bar{s}$, $r$ and $t_i$ are eigenvalues of $\mathbb{D}$, $\mathbb{L}_1{}^1$,
$\bar{\mathbb{L}}^{\dot{1}}{}_{\dot{1}}$, $\mathbb{R}$ and $\mathbb{T}_i{}^i$, respectively.
These numbers obey obvious inequalities, $n_1 \leq n_2 \leq n_3 \leq n_4 \geq n_5 \geq n_6 \geq
n_7$, $L \geq n_6 - n_7$ and $L \geq n_5 - n_6$.

\section{Young tableaux for FBBFFBBF grading}
\label{su2Dynkin}

Considerations presented in the main body of the paper can be applied in a straightforward
manner to the Kac-Dynkin diagram in the FBBFFBBF grading, i.e., with the compact central
$su(2)$ node. The elementary Young tableaux read in this case
\ba
\unitlength0.4cm
\begin{picture}(2.3,1.2)
\linethickness{0.4mm}
\put(1,0){\framebox(1,1){$1$}}
\end{picture}
{}_{u}
\!\!\!&=&\!\!\!
\left( x^- \right)^L
\frac{
\widehat{Q}^{(1)} \left( u^+ \right)}{\widehat{Q}^{(1)} \left( u^- \right)}
{\rm e}^{\ft12
\left[
\sigma^{(4)}_+ (u^-) - \sigma^{(4)}_- (u^-) + \sigma^{(4)}_- (u^+) - \sigma^{(4)}_+ (u^+) + 2 \Theta (u^+)
+ \sigma_0^{(3)} (u^-) - \sigma_0^{(3)} (u^+)
\right]}
\, , \\
\unitlength0.4cm
\begin{picture}(2.3,1)
\linethickness{0.4mm}
\put(1,0){\framebox(1,1){$2$}}
\end{picture}
{}_{u}
\!\!\!&=&\!\!\!
\left( x^- \right)^L
\frac{
\widehat{Q}^{(1)} \left( u^+ \right)}{\widehat{Q}^{(1)} \left( u^- \right)}
\frac{
Q^{(2)} \left( u - i \right)}{Q^{(2)} \left( u \right)}
{\rm e}^{\ft12
\left[
\sigma^{(4)}_- (u^+) - \sigma^{(4)}_+ (u^+) + 2 \Theta (u^+)
+ \sigma_0^{(3)} (u^-) - \sigma_0^{(3)} (u^+)
\right]}
\, , \nonumber\\
\unitlength0.4cm
\begin{picture}(2.3,1)
\linethickness{0.4mm}
\put(1,0){\framebox(1,1){$3$}}
\end{picture}
{}_{u}
\!\!\!&=&\!\!\!
\left( x^- \right)^L
\frac{
Q^{(2)} \left( u + i \right)}{Q^{(2)} \left( u \right)}
\frac{
\widehat{Q}^{(3)} \left( u^- \right)}{\widehat{Q}^{(3)} \left( u^+ \right)}
{\rm e}^{\ft12
\left[
\sigma^{(4)}_- (u^+) - \sigma^{(4)}_+ (u^+) + 2 \Theta (u^+)
+ \sigma_0^{(1)} (u^+) - \sigma_0^{(1)} (u^-)
\right]}
\, , \nonumber\\
\unitlength0.4cm
\begin{picture}(2.3,1)
\linethickness{0.4mm}
\put(1,0){\framebox(1,1){$4$}}
\end{picture}
{}_{u}
\!\!\!&=&\!\!\!
\left( x^- \right)^L
\frac{
\widehat{Q}^{(3)} \left( u^- \right)}{\widehat{Q}^{(3)} \left( u^+ \right)}
\frac{
Q^{(4)} \left( u + i \right)}{Q^{(4)} \left( u \right)}
{\rm e}^{\ft12
\left[
2 \Theta (u^+) + \sigma_0^{(1)} (u^+) - \sigma_0^{(1)} (u^-)
\right]}
\, , \nonumber\\
\unitlength0.4cm
\begin{picture}(2.3,1)
\linethickness{0.4mm}
\put(1,0){\framebox(1,1){$5$}}
\end{picture}
{}_{u}
\!\!\!&=&\!\!\!
\left( x^+ \right)^L
\frac{
\widehat{Q}^{(5)} \left( u^+ \right)}{\widehat{Q}^{(5)} \left( u^- \right)}
\frac{
Q^{(4)} \left( u - i \right)}{Q^{(4)} \left( u \right)}
{\rm e}^{\ft12
\left[
2 \Theta (u^-) + \sigma_0^{(7)} (u^-) - \sigma_0^{(7)} (u^+)
\right]}
\, , \nonumber\\
\unitlength0.4cm
\begin{picture}(2.3,1)
\linethickness{0.4mm}
\put(1,0){\framebox(1,1){$6$}}
\end{picture}
{}_{u}
\!\!\!&=&\!\!\!
\left( x^+ \right)^L
\frac{
Q^{(6)} \left( u - i \right)}{Q^{(6)} \left( u \right)}
\frac{
\widehat{Q}^{(5)} \left( u^+ \right)}{\widehat{Q}^{(5)} \left( u^- \right)}
{\rm e}^{\ft12
\left[
\sigma^{(4)}_+ (u^-) - \sigma^{(4)}_- (u^-) + 2 \Theta (u^-)
+ \sigma_0^{(7)} (u^-) - \sigma_0^{(7)} (u^+)
\right]}
\, , \nonumber\\
\unitlength0.4cm
\begin{picture}(2.3,1)
\linethickness{0.4mm}
\put(1,0){\framebox(1,1){$7$}}
\end{picture}
{}_{u}
\!\!\!&=&\!\!\!
\left( x^+ \right)^L
\frac{
\widehat{Q}^{(7)} \left( u^- \right)}{\widehat{Q}^{(7)} \left( u^+ \right)}
\frac{
Q^{(6)} \left( u + i \right)}{Q^{(6)} \left( u \right)}
{\rm e}^{\ft12
\left[
\sigma^{(4)}_+ (u^-) - \sigma^{(4)}_- (u^-) + 2 \Theta (u^-)
+ \sigma_0^{(5)} (u^+) - \sigma_0^{(5)} (u^-)
\right]}
\, , \nonumber\\
\unitlength0.4cm
\begin{picture}(2.3,1.2)
\linethickness{0.4mm}
\put(1,0){\framebox(1,1){$8$}}
\end{picture}
{}_{u}
\!\!\!&=&\!\!\!
\left( x^+ \right)^L
\frac{
\widehat{Q}^{(7)} \left( u^- \right)}{\widehat{Q}^{(7)} \left( u^+ \right)}
{\rm e}^{\ft12
\left[
\sigma^{(4)}_- (u^+) - \sigma^{(4)}_+ (u^+)
+ \sigma^{(4)}_+ (u^-) - \sigma^{(4)}_- (u^-) + 2 \Theta (u^-)
+ \sigma_0^{(5)} (u^+) - \sigma_0^{(5)} (u^-)
\right]}
\, .
\ea
The grading of each node is inverted compared to the Kac-Dynkin diagram in Fig.\
\ref{Dynkin}, i.e., $\bar{1} = \bar{4} = \bar{5} = \bar{8} = 1$ and $\bar{2} =
\bar{3} = \bar{6} = \bar{7} = 0$. The fused transfer matrices are built along
the procedures outlined in Sections \ref{AnalyticBA} and \ref{Tsystem}.

\section{Perturbative solution of $sl(2)$ Baxter equation}
\label{SolutBaxter}

For the reader's convenience, we present here a sample Mathematica routine to
compute anomalous dimensions in the $sl(2)$ sector to two loop order from the
Baxter equation
\be
(x^+)^L {\rm e}^{\Delta_+ (u^+)} Q^{(4)} (u + i)
+
(x^-)^L {\rm e}^{\Delta_- (u^-)} Q^{(4)} (u - i)
=
t (u) Q^{(4)} (u)
\, .
\ee
The perturbative expansion for Baxter polynomials and transfer matrix
\be
Q^{(4)} (u) = Q^{(4)}_0 (u) + g^2 Q^{(4)}_1 (u) + \mathcal{O} (g^4)
\, , \qquad
t (u) = t_0 (u) + g^2 t_1 (u) + \mathcal{O} (g^4)
\, ,
\ee
and corresponding expansion of the dressing factors is implemented in the
following module:
{\footnotesize
\begin{verbatim}
Spectrum[spin_, length_] := Module[{J, L, BaxEq0, BaxEq1, k, n, Q40, Q41, QQ0, QQ1},
J = spin; L = length;
Q40[u_] = u^J + Sum[a[k] u^k, {k, 0, J - 1}];
Q41[u_] = Sum[b[k] u^k, {k, 0, J - 1}];
t0[u_] = 2 u^L - ((J + L/2)(J + L/2 - 1) + L/4 ) u^(L - 2) + Sum[q0[k] u^(L - k), {k, 3, L}];
t1[u_] = Sum[q1[k] u^(L - k), {k, 1, L}];
BaxEq0 := Q40[u] t0[u] - (u + I/2)^L Q40[u + I] - (u - I/2)^L Q40[u - I];
BaxEq1 := (Q41[u] t0[u] + Q40[u] t1[u])
        - (- (L/4)(u + I/2)^(L - 2) Q40[u + I] + (u + I/2)^L Q41[u + I]
        - (u + I/2)^(L - 1) Q40[u + I] Q40'[I/2]/(2 Q40[I/2]))
        - (- (L/4)(u - I/2)^(L - 2) Q40[u - I] + (u - I/2)^L Q41[u - I]
        - (u - I/2)^(L - 1) Q40[u - I] Q40'[-I/2]/(2 Q40[-I/2]));
QQ0 = Q40[u]; QQ1 = Q41[u];
BaxEq0 = Collect[BaxEq0, u]; BaxEq1 = Collect[BaxEq1, u];
Do[Sol0 = Solve[Coefficient[BaxEq0, u, L + J - 3 - (k - 1)] == 0, a[J - k]];
          BaxEq0 = Simplify[BaxEq0/.Sol0[[1]]]; QQ0 = Simplify[QQ0/.Sol0[[1]]];
          BaxEq1 = Simplify[BaxEq1/.Sol0[[1]]], {k, 1, J}];
SolutionQQex = NSolve[Table[Coefficient[BaxEq0, u, k] == 0, {k, 0, L - 3}],
                      Table[q0[k + 3], {k, 0, L - 3}], 100];
Do[SolQ11 = Solve[Coefficient[BaxEq1, u, L + J - k] == 0, q1[k]];
   BaxEq1 = Simplify[BaxEq1/.SolQ11[[1]]];
   QQ1 = Simplify[QQ1/.SolQ11[[1]]], {k, 1, L}];
Do[Sol1 = Solve[Coefficient[BaxEq1, u, J - k] == 0, b[J - k]];
   BaxEq1 = Simplify[BaxEq1 /. Sol1[[1]]];
   QQ1 = Simplify[QQ1/.Sol1[[1]]], {k, 1, J}];
En0 = Simplify[Factor[(I*(D[QQ0, u]/(2 QQ0))/.u -> I/2)
                    - (I*(D[QQ0, u]/(2 QQ0))/.u -> -I/2) ]];
En1 = Simplify[Factor[(I*(D[QQ0, u]^3/(8 QQ0^3)
                          - 3 D[QQ0, u] D[QQ0, {u, 2}]/(16 QQ0^2)
                          + D[QQ0, {u, 3}]/(16 QQ0) + D[QQ1, u]/(2 QQ0)
                          - D[QQ0, u] QQ1/(2 QQ0^2))/.u -> I/2)
                    - (I*(D[QQ0, u]^3/(8 QQ0^3)
                          - 3 D[QQ0, u] D[QQ0, {u, 2}]/(16 QQ0^2)
                          + D[QQ0, {u, 3}]/(16 QQ0) + D[QQ1, u]/(2 QQ0)
                          - D[QQ0, u] QQ1/(2 QQ0^2))/.u -> -I/2) ]];
LogTheta0 = Log[Factor[(QQ0 /. u -> I/2)/(QQ0/.u -> -I/2)]];
Dat = {{Gamma0 -> Rationalize[ En0/.SolutionQQex[[1]]],
        Gamma1 -> Rationalize[En1/.SolutionQQex[[1]]],
        logTheta -> Chop[N[LogTheta0/.SolutionQQex[[1]]]]}};
Do[Dat = Join[Dat, {{Gamma0 -> Rationalize[En0/.SolutionQQex[[n]]],
         Gamma1 -> Rationalize[En1/.SolutionQQex[[n]]],
         logTheta -> Chop[N[LogTheta0/.SolutionQQex[[n]]]]}}],
         {n, 2, Length[SolutionQQex]}];
Dat]
\end{verbatim}
}
\noindent The output is presented in the form \{{\tt\footnotesize Gamma0}$\to\gamma_0$,
{\tt\footnotesize Gamma1}$\to\gamma_1$, {\tt\footnotesize logTheta}$\to \ln \theta$\},
with $\gamma_0$ and $\gamma_1$ being the numerical values of anomalous dimensions at
one and two loops, respectively, and $\log\theta$ is the value of the corresponding
quasimomentum $\theta$ of the state. Though we used the explicit form of the conserved
charge {\tt\footnotesize q0[2]} in terms of eigenvalues of the bare quadratic Casimir
$\mathbb{C}_2 (g = 0) = - (N + \ft12 L)(N + \ft12 L - 1) - \ft14 L$ of the $sl(2)$
algebra where $n_4 = N$ is the number of excitations and $L$ is the length of chain,
cf.\ Eq.\ \re{sl21quadraticCasimir}, it is not a necessary input as its value can be
determined from the Baxter equation itself as a coefficient in front of the highest
power of the spectral parameter $u^{L + N - 2}$. Generalization to higher loop orders
is straightforward and analyses were presented elsewhere \cite{Bel06}. Baxter
equations for higher rank algebras can be treated analogously with results for
$sl(2|1)$ long-range spin chain presented in Section \ref{sl21diagonalization}.



\end{document}